\documentstyle[epic,eepic]{article}

\newtheorem{de}{Definition}
\newtheorem{pr}{Proposition}

\newtheorem{th}{Theorem}
\def\ex{\times}
\def\oo{\circ}
\newtheorem{alg}{Algorithm}
\def\zz{\mbox{\bf Z}}
\def\cc{\mbox{\bf C}}

\def\kkr{{\sc kkr}}

\begin{document}

\title{
On the combinatorics of row and corner transfer matrices of
the $A_{n-1}^{(1)}$ restricted face models
}

\author{Srinandan Dasmahapatra\\
       \em Department of Mathematics,\\
       \em City University,\\
       \em London, EC1V 0HB, UK}

\date{December 1995}

\maketitle

\begin{abstract}
We establish a weight-preserving bijection
between the index sets of
the spectral data of row-to-row and corner transfer matrices for
$U_q\widehat{sl(n)}$ restricted interaction round a face (IRF) models.
The evaluation of momenta by adding Takahashi integers in the
spin chain language is shown to directly correspond to
the computation of the energy of a path on the
weight lattice in the two-dimensional model.
As a consequence we derive fermionic forms of
polynomial analogues of branching functions
for the cosets
${(A^{(1)}_{n-1})_{\ell -1}\otimes (A^{(1)}_{n-1})_{1}}
\over (A^{(1)}_{n-1})_{\ell}$, and establish a bosonic-fermionic
polynomial identity.
\end{abstract}

\section{Introduction and Summary}

The basic object of interest in the statistical mechanics
of a large number of classically
interacting degrees of freedom in $d$ Euclidean space-time dimensions
is the partition function, which is the generating function of
all correlation functions.
The calculation of the partition function requires the
(Boltzmann) weighted sum over all configurations of the
system, and can be generated by a {\em transfer matrix}.
{}From the transfer matrix, it is possible
to extract the definition of an interacting
hamiltonian for a quantum system in $(d-1)$ spatial dimensions.
Even though the physical interpretation in the two cases is often very
different, techniques developed for investigation in one framework can
be adapted to extract useful information in another.  For $d=2$, there
have been very striking recent advances in the study of exactly solvable
models which do precisely this.

The method of introducing the {\em corner transfer matrix} (CTM)
\cite{BaxterBook} for the calculation of order parameters
\cite{ABF,DJMO,JKMO,JMO} has spawned
the development of quantum affine Lie algebraic tools for the study
of correlation functions, after the realization that the CTM density
is closely related to the derivation in the affine algebra \cite{FM,five}.
These have then been applied to the study of
$XXZ$ (for $U_q(\widehat{sl_2})$) and higher rank {\em infinitely long}
quantum chains.  Among other results, this method has provided a
mathematical (representation theoretic) grounding to the notion
of a ``quasi-particle'' \cite{FT} in this model \cite{five}.

The concept of a quasi-particle is a key physical ingredient in
the intuition built into the search for collective modes which
appear as summands in the additive decomposition of the
energy and
momentum of the many-body system
in the thermodynamic limit.  In the context of the
ferromagnetic $XXX$ Heisenberg chain, a magnon or spin-wave
describes a linear combination of states that have this additive
character, and is called an elementary excitation.  For the
anti-ferromagnetic chain, the Bethe ansatz approach requires
filling the ``Fermi sea''\footnote{This terminology is borrowed
from the description of many-fermion systems.}
with ``pseudo-particles'' (magnonic or spin-wave states)
and looking for ``dressed'' particle
and ``hole'' excitations (filling an unfilled state, or evacuating
an occupied one), above the Fermi sea.
The concept of ``spinons'' that emerged from this
picture \cite{FT} was put into the language of representation theory
in \cite{five}, in terms of the span of a momentum-indexed
set of eigenstates related by non-Abelian symmetries of the
hamiltonian.

There has been another set of developments in the past few years
which have conjectured \cite{KM,DKMM} (and proven in some cases
\cite{proofs,bijective})
that certain generating functions of momenta of
fermionic quasi-particle
excitations above a ground state in quantum chains coincide with the
low-temperature expansions of the order parameter in the associated
$2$-dimensional models.  In the case of quantum spin chains, the
eigenvalues of the {\em row transfer matrix} (RTM) were studied.
It was the observation that these order
parameters of the $2$-dimensional models were, in fact,
branching functions
of cosets of affine Lie algebras that led to the development
and implementation of ideas from representation theory.
It seems equally likely that the calculation
of the spectrum of the RTM
approach may be similarly recast in more mathematically
precise terms.

Both the CTM and RTM calculations under consideration
are essentially combinatoric in nature, {\em i.e.}
both involving {\em counting} objects in a certain class.
Namely, in the CTM approach, what is evaluated is a weighted sum
of a set of paths on the
weight lattice of the appropriate quantum affine algebra
which index the eigenvectors of the CTM, whereas in the
RTM calculations, the
weighted sum performed is over the configurations
of strings and holes which characterize
solutions to the Bethe equations, on which the
eigenvalues of the RTM depend.
In this paper, we pursue the idea that since both these combinatorial
approaches give the same answer, there must be a bijection between the
sets being counted.  In a number of cases \cite{bijective}
bijective proofs have been provided for polynomial analogues
of these branching function identities.  However, they do not address
the issue of mapping the objects counted in the two different
cases into each other.  It was observed in
\cite{DF0,DF1} however, that one could write down an explicit
bijection between these two sets of objects, based on
an algorithmic construction given in a remarkable paper \cite{KKR}.
The face models that were studied were related to the cosets
${(A^{(1)}_{n-1})_{\ell -1}\otimes
(A^{(1)}_{n-1})_{1}} \over (A^{(1)}_{n-1})_{\ell}$,
for the cases arbitrary $n$, $\ell=2$ and arbitrary $\ell$, $n=2$.
The present paper extends this idea to arbitrary $(n,\ell)$.
The two solutions, for the CTM and RTM approaches, that exist in
the literature for these models are those of the order parameters
in \cite{JMO} and the spectrum and thermodynamics in \cite{BRtwo}.

\subsection{Finite size approximations}

The bijection \cite{KKR} of this paper and \cite{DF1} is established
for finite chains, where we do not have the full affine symmetry.
This may be viewed as an approximation of the infinite lattice
case in a certain sense to be explained below.
Reference \cite{KKR} established a bijection between the ``rigged
configurations'' which parametrize Bethe vectors, which are highest
weight vectors of $gl(n)$ (for $gl(n)$ invariant models) and the
well-known index set of highest weight states of $gl(n)$, namely,
Young tableaux \cite{FH}.
These ``rigged configurations'' were based on the `string
hypothesis' \cite{Bethe} form of the solutions of the Bethe
equations, the counting method of \cite{Tak}
generalized for the higher rank models \cite{nested}.
By Schur-Weyl duality, the cardinality
of these sets is the dimension of the irreducible
representation of $\mbox{\sf S}_L$, the symmetric group on $L$
letters ($L$ is the size of the system), labelled by the
partition $\lambda$ of $L$.  This is called the {\em Kostka number},
$K_{\lambda \mu}$ the number of tableaux of shape
$\lambda$ and weight $\mu$,
where $\mu=1^L$ and $\lambda\vdash L$. It can be expressed
in terms of a vector
partition function $\pi(\mu)$, defined by
$$\prod_{\alpha\in R^+}{1\over 1-e^{\alpha}}=\sum_{\mu}\pi(\mu)e^{\mu},$$
which counts the number of ways of writing a weight $\mu$
as a sum of positive roots $\alpha\in R^+$.
The Kostant multiplicity formula\footnote{This is in fact equivalent
to the Weyl character formula.}
\begin{equation}
n_\mu(\Gamma_\lambda)=\sum_{w\in \bar{W}} (-1)^{l(w)}
\pi\bigl(w(\lambda+\bar{\rho})-(\mu+\bar{\rho})\bigr),
\end{equation}
gives the number of times the weight $\mu$ occurs in the irreducible
representation $\Gamma_{\lambda}$ labelled by the highest weight
$\lambda$.  Here $\bar{\rho}$ is half the sum of the positive roots,
$\bar{W}$ is the Weyl group for the root system of the Lie algebra,
and and any $w\in\bar{W}$ can be expressed as a product
of $l(w)$ elementary reflections, where $l(w)$ is called the length of
the word in $\bar{W}$.
For the $A_{n-1}$ root system,
$n_\mu(\Gamma_{\lambda})=K_{\lambda\mu}$ and $\bar{W}$ is isomorphic
to the symmetric group.

The bijection of reference \cite{KKR} established the ``completeness''
of Bethe ansatz states by demonstrating that the number of Bethe states
was indeed equal to the Kostka number.  In other words, this established
that the ``bare'' or ``pseudo-particle'' description of the spectrum,
as an assembly of spin-wave-like states
accounted for all the states in any finite system,
\footnote{It should be mentioned, however, that the Bethe states are
only the highest weight states, and the rest of the states are obtained
by the appropriate ``lowering'' operators.}
so that the
``dressed'' quasi-particle states above the ground state would
guarantee the {\em completeness of the particle picture} in the
emergent theory in the thermodynamic limit.  The spectral decomposition
of all
correlation functions in this theory would involve
summing over these quasi-particle states.

The vector partition function may be refined as follows:
$$\prod_{\alpha\in R^+}{1\over 1-qe(\alpha)}=\sum_\mu \pi_q(\mu)e(\mu),$$
where the coefficient of $q^k$ in $\pi_q(\mu)$ counts the number
of ways a weight $\mu$ can be expressed as a sum of $k$ positive
roots $\alpha\in R^+$.  The $q$-analogue of the Kostant
multiplicity formula,

\begin{equation}
K_{\lambda \mu}(q)=\sum_{w\in \bar{W}} \varepsilon(w)
\pi_q\bigl(w(\lambda+\bar{\rho})-(\mu+\bar{\rho})\bigr),
\label{kf-bosonic}
\end{equation}
is the Kostka-Foulkes polynomial.\footnote{This is the special case
(for the root system $A_{n-1}$) of the (normalized) Kazhdan-Lusztig
polynomials for the affine Weyl group.}  $K_{\lambda\mu}(q)$ are the
entries of the transition matrix between two bases in the ring of
symmetric functions in a denumerably infinite set of independent
variables
$X=x_1,x_2,\ldots$, namely the {\em Schur function} $s_{\lambda}(X)$,
and $P_{\lambda}(X;q)$ called
the {\em Hall-Littlewood function}
\begin{equation}
s_{\lambda}(X)=\sum_{\mu}K_{\lambda \mu}(q)P_{\lambda}(X;q).
\end{equation}
Explicitly, for $x_{n+i}=0$ for $i=1,2,\ldots$ in $X$ and
$\lambda=(\lambda_1,\ldots,\lambda_n)$ a partition of length $\leq n$.
$$P_{\lambda}(X;q)={1\over {v_{\lambda}(q)}}\sum_{w\in\mbox{\sf S}_n}
w\Bigl(x_1^{\lambda_1}\cdots x_n^{\lambda_n}\prod_{i<j}{{x_i-qx_j}\over
{x_i-x_j}}\Bigr)$$
where
$$v_{\lambda}=\prod_{j\geq 1}\prod_{i=1}^{m_j}(1+q+\cdots+q^{i-1})$$
for $m_i=|\{j|\lambda_j=i\}|$, the number of rows of $\lambda$ of length
$i$.  For $q=0$, the Hall-Littlewood
function reduces to
$$S_{\lambda}(X)=det(x_i^{\lambda_j+n-j})/det(x_i^{n-j}),$$
and it
reduces to the {\em monomial symmetric function} $m_{\lambda}(X)$
at $q=1$.  This implies that the Kostka numbers (Kostant mutiplicities)
are also entries of the transition matrix between Schur and monomial
symmetric functions.  Both $S_{\lambda}$ and $m_{\lambda}$ are bases
for the ring of symmetric functions $\Lambda$, so the
$P_{\lambda}(X;q)$ interpolate between them.

The bijection of \cite{KKR} was extended in \cite{KR} for semi-standard
tableau, and it was also shown to encode the statistic {\em charge}
of a tableau, $c(T)$, a natural number-valued function of tableaux
of shape $\lambda$ and weight $\mu$
introduced in \cite{L-Sch}, so that the $c$-weighted sum over such
tableaux gives the Kostka-Foulkes polynomial
\begin{equation}
K_{\lambda \mu}(q)=\sum_{\begin{array}{c}
\scriptstyle T\\
\mbox{\small sh}\scriptstyle T = \lambda \\
\mbox{\small wt}\scriptstyle T = \mu
\end{array}}
q^{c(T)}.
\end{equation}
Namely, it was shown that a ``bare'' momentum generating function
defined for a length $L$ set of configurations $\{\nu\}$ of
$s_j^{(a)}$ strings and $h_j^{(a)}$ holes
of length $j\geq 1$ and colour $(a)=1,\ldots,n-1$
coincides with the
Kostka-Foulkes polynomial \cite{KR,K-JGP}:
\begin{equation}
\sum_{\{T\}} q^{c(T)}=\sum_{\{\nu\}}
q^{{1\over 2}L(L-1-\sum_j s_j^{(1)})}
\prod_{a=1}^{n-1}\prod_{j\geq1}
q^{-{1\over 2}s_j^{(a)}h_j^{(a)}} \biggl[
\begin{array}{c}
s_j^{(a)}+h_j^{(a)}\\
s_j^{(a)}
\end{array}\biggr],
\label{kf-fermionic}
\end{equation}
where $\nu=\rho(T)$ and $T=\tau(\nu)$ by the bijection of ref \cite{KKR,KR}
and
$$\biggl[
\begin{array}{c}
a+b\\
b
\end{array}\biggr]={[a+b]!\over{[a]! [b]!}},\,\,[k]!=[k][k-1]\cdots[1],
\, \, [k]:=\prod_{i=1}^{k}(1-q^i).$$
The power of $q$ in the expansion on the right hand side is related
to the sum of the Takahashi integers indexing each state in the model,
which gives the momentum of the Bethe state.
Note that while (\ref{kf-fermionic}) is a sum of manifestly positive
terms, (\ref{kf-bosonic}) has terms which alternate in sign.
In view of what is to follow,
equations (\ref{kf-fermionic}) and (\ref{kf-bosonic}) can be called
a bosonic-fermionic polynomial identity for Kostka-Foulkes
polynomials.

The appropriate objects for the {\em restricted}
models we are interested in in this paper, are a certain subset of
irreducible representations (tilting modules) of $U_q sl_n$ for
$q=e^{{2\pi \imath}\over {\ell+n}}$ (p.261, \cite{chari}.)
In \cite{wenzl} the Schur-Weyl dual, the
irreducible quotient of the Hecke algebras ${\sf H}_L(q)$,
is given in terms of a path description.
\footnote{For similar constructions, see also \cite{GW,GN},
and \cite{mathieu} for applications to modular representation
theory of $\mbox{\sf S}_L$ over $\bar{F}_{\ell+n}$.}
This path description is identical to that given in \cite{JMO}
(see \cite{GN} for a discussion and proof), and in section 2 of this
paper.

In \cite{JMMO} it was shown that the $L$-fold tensor product of the
$\ell^{\mbox{\tiny th}}$ symmetric tensor representation of
$U_q sl_n(\cc)$ ``approximates'' the path realization of the crystal base
of level $\ell$ $U_q \widehat{sl_n}$ modules.
In \cite{MM,JMMO} it was proved that the path construction
inspired by CTM calculations is convenient to define the crystal basis
of a $U_q\widehat{sl_n}$ highest weight module.\footnote{This is developed
further in refs. \cite{KKMMNN} where the theory of perfect and affine crystals
is introduced.} In \cite{JMMO}, an important observation regarding
finite-sized approximations to the crystal graph was made.
It was shown that the crystal structure of
a finitized path of length $L$ as inherited from the infintely long
path is isomorphic to that induced from an $L+1$-fold tensor product
of irreducible finite dimensional representations into which the finite
path may be embedded.

A {\em finitized path}
of size $L$ is one which looks like a ground state path from at least
step $L+1$ of the path onwards. For $\Lambda_i, i=0,\ldots,n-1$
the fundamental
weights of $U_q\widehat{sl_n}$, $\delta$ the null root, and
$P=\oplus_{i=0}^{n-1}\zz \Lambda_i \oplus \zz\delta$,
the weight lattice, a finitized path is a sequence
$(\mu_k)_{k\geq 0}\, \mu_k\in P$ (and other auxiliary conditions
defined in Section 2) such that $\mu_{L+i}=\Lambda'+\Lambda_{i+j-1}$
for some $j=0,1,\ldots,n-1$, $\Lambda'\in P$ and $i>1$.  Note the
cyclic property of
the Dynkin diagram $\Lambda_{i+kn}=\Lambda_i$ for all integer $k$.

Thus, the multiplicities
of weights in the module of the affine algebra may be approximated
by a consideration of corresponding Kostant formulas for the finite
dimensional modules of $U_q sl_n$.
In fact, such an analog of Kostant's $q$-multiplicity
formula has been considered in \cite{JMO} where the weight (the
exponent of $q$) is given by the `energy function' obtained by
taking the $q=0$ limit of the Boltzmann weights of these restricted
models.  This is used to evaluate the eigenvalue of the
CTM at $q=0$, when the paths diagonalize the CTM hamiltonian.
It was established in \cite{JMO} that for a finitized path, or
sequence of weights $\mu=(\mu_0,\mu_1,\ldots,\mu_L,\mu_{L+1})$,
such that each $\mu_j$ is of the form $\sum_{i=0}^{n-1} m_i
\Lambda_i$ and the $m_i$s are non-negative integers satisfying
$\sum_i m_i=\ell$, and $\mu_{i+1}-\mu_i=\Lambda_j-\Lambda_{j+1}$
for some $j$, the following is true.
For a section of a path $(\mu_i,\mu_{i+1}\mu_{i+2})$ where
$\mu_j\in P$,
one can define a local energy function $H$, from the $q=0$ (low
temperature) limit of the Boltzmann weights:
$$H(\Lambda,\Lambda+\hat{\jmath},\Lambda+\hat{\jmath}+\hat{\imath})
=\theta(\imath-\jmath),$$
where $\hat{\jmath}:=(\Lambda_{j+1}-\Lambda_j)$
for $\jmath=0,1,\ldots,n-1$ and
$\theta$ is the step function given by
$$\theta(z) = \left\{
\begin{array}{ll}
0 \qquad & (z  >   0) \\
1        & (z \leq 0).
\end{array}
\right.$$

In order to compute the one-point functions of the restricted
face models of $A^{(1)}_{n-1}$,
the following weighted sum over the set of paths $\mu$ was
calculated in \cite{JMO}:
\begin{equation}
X_L(a,b,c;q)=
\displaystyle
\sum_{\begin{array}{c}
\{\scriptstyle \mu | {\mu_0=a, \mu_L=b,  \mu_{L+1}=c}\}
\end{array}}
\! \! \! \! \! \! q^{\sum_{k=0}^{L-1}(k+1)H(\mu_k,\mu_{k+1},\mu_{k+2})},
\end{equation}
where the steps $\mu_k$ of the path belong to the set
$$P^+(n;\ell)=\{ \Lambda=\sum_{i=0}^{n-1} m_i \Lambda_i |
m_i \in \zz, \, m_i \geq 0, \, \sum_{i=0}^{n-1} m_i = \ell \}$$
of dominant integral weights.
For
$$b=\xi+\sigma^{L}(\eta), \, \, \mbox{for}\, \, \xi\in P^+(n;\ell-1),
\eta\in P^+(n;1),\quad
c=b+\Lambda_{k+1}-\Lambda_k,$$ it was found that
\begin{equation}
X_L(a,b,c;q)=
\sum_{w\in W} (-1)^{l(w)} q^{{1\over 2}|\xi+\rho-w(a+\rho)|^2+c_L}
{[L]!\over{\prod_{i=0}^{n-1}[\alpha_i]!}},
\label{jmo-bosonic}
\end{equation}
where $\{\alpha_i\}$ satisfies the relations
$$\sum_{i=0}^{n-1}\alpha_i = L,\, \, \mbox{and} \, \,
\sum_{i=0}^{n-1}\alpha_i (\Lambda_{i+1}-\Lambda_i)\equiv b+\rho -w(a+\rho)\,
\mbox{mod}\, \delta,$$
$\rho=\sum_{i=0}^{n-1}\Lambda_i=\bar{\rho}+n\Lambda_0$,
$W$ is the affine Weyl group, $l(w)$
is the length function, and
\begin{equation}
c_{L}(j)={1\over 2n}(L-j)(L+n-j).
\end{equation}
The inner product on the weight lattice used above is given by
$$\langle\Lambda_i,\Lambda_j\rangle=\mbox{min}(i,j)-{ij\over n}=:G^{-1}_{ij},
\, \, \langle\delta,\delta\rangle=0 \, \,\mbox{and}
\, \, \langle \delta, \Lambda_i \rangle=1.$$

By the isomorphism
between the crystal graph of the highest weight representations
of $U_q \widehat{sl_n}$ and the
appropriate construction of the action of the
Kashiwara endomorphisms on the CTM paths \cite{JMMO}, the
$L\rightarrow\infty$ limit of these polynomials gives the branching
functions of the cosets
${(A^{(1)}_{n-1})_{\ell -1}\otimes (A^{(1)}_{n-1})_{1}}
\over (A^{(1)}_{n-1})_{\ell}$:
$$\chi_{\xi}(z_1,\ldots,z_n;q)\chi_{\eta}(z_1,\ldots,z_n;q)=
\sum_a
b_{\xi\eta\,a}(q) \chi_a(z_1,\ldots,z_n;q),$$
in the principal specialization ($z_i=q^{1/n}$)
where $\chi_{\alpha}$ are the characters of the irreducible modules
with highest weight $\alpha$.  (For expressions for $\chi_{a}$
and $b_{abc}$, see \cite{JMO}.)  Explicitly, we have
\begin{equation}
\lim_{L\rightarrow\infty} q^{\phi_L}X_L(a,b,c;q)
=b_{\xi\eta\,a}(q)
\end{equation}
where
$\phi_L=L\bigl((L/n)-1\bigr)/2+\gamma(\xi,\eta,a)$,
$$\gamma(\xi,\eta,a):={|\xi+\rho|^2 \over {2 (\ell+n-1)}}+
{|\eta+\rho|^2 \over {2 (n+1)}}-
{|a+\rho|^2 \over {2(\ell+n)}}-
{|\rho|^2 \over {2n}}.$$

The presence of alternating
signs in (\ref{jmo-bosonic}) which carries over to the branching
functions in the limit,
is explained as the elimination of states
from the $(n-1)$ free bosonic Fock space realization of the Verma
modules \cite{fat-lyk,feig-fuks},
and is hence called a {\em bosonic} form of the branching
function.   In this paper, we prove
the equality of the polynomials $X_L(\ell\Lambda_0,\ell\Lambda_0,
(\ell-1)\Lambda_0+\Lambda_1)(q)$ and the sum (\ref{myform})
below, whose terms have manifestly
positive coefficients, called a {\em fermionic} form. This
polynomial was first written down in \cite{das} and its equality
with (\ref{jmo-bosonic}) conjectured in \cite{pw}, in the spirit
of \cite{melzer}.

The proof is based on two observations.  It is well known
\cite{Macdonald} that for standard tableaux, the statistics charge
(mentioned earlier) and ${\bf p}$ introduced in \cite{thomas}
are equidistributed.  It was observed in \cite{DF1} that ${\bf p}$
evaluated on a standard tableau designed to encode a path
{\em is  identical to} the energy function
$\sum_{j=0}^{L-1}(j+1)H(\mu_j,\mu_{j+1},\mu_{j+2})$ evaluated on the
same path.  The second
observation is the property of the bijection
of \cite{KKR} that it respects the level of the integral weights
in a certain sense.  Namely, for a tableau encoding a restricted path
of level $\ell$, it can be proved that the maximum length of string
allowed in the bijectively associated configuration of strings and holes
is $\ell$, as was conjectured in \cite{BRtwo} after numerical
investigation of the solutions of Bethe's equations.  This immediately
leads to the proof of the polynomial identity,

\begin{th}[bosonic-fermionic polynomial identity]

\begin{equation}
q^{-{1\over 2}L({L\over n}-1)}X_{L-1}=
\sum_{\{\nu\}}
\prod_{a=1}^{n-1}\prod_{j=1}^{\ell}q^{{1\over 2}({\bf h}|{\bf h})}
\biggl[
\begin{array}{c}
h_j^{(a)}+\sum_b \sum_k G^{-1}_{ab} C_{jk}h_k^{(b)}\\
h_j^{(a)}
\end{array}\biggr],
\end{equation}
where
$$({\bf h}|{\bf h})=\sum_{ab\,;\,jk}
G^{-1}_{ab} C_{jk}h_j^{(a)}h_k^{(b)},$$
and $C_{jk}=(2\delta_{jk}-\delta_{j (k-1)}-\delta_{j (k+1)})$.
\end{th}

The paper is organized as follows.  We give the basic definitions
of the model and describe the CTM paths in section 2, and how to
encode them as standard tableaux.  In section 3, we define functionals
on the set of paths and tableaux, and demonstrate the relationship
between them.  In section 4, we introduce the `rigged configurations'
and in sections 5 and 6, we describe the bijection of \cite{KKR},
and another version \cite{Han} of the
same, which is cast in the familiar language of strings and holes,
and has a pictorial character.   In section 7, we evaluate the
momenta of the Bethe states in terms of a sum of Takahashi
integers, and prove that this gives the charge or {\bf p}
of the corresponding tableau.  Here we give the proofs omitted in
\cite{KR,K-JGP}.  This leads immediately to the proof of the
equidistribution of energies of CTM paths and momenta of Bethe states.
This is stated in section 8, after which we end with a small
discussion.

\section{The state space, tableaux and paths.}

In RSOS models, the `spin' degrees of freedom assigned at each node
of the lattice
take on any values from the set (state space) $P^+(n;\ell)$, whose
elements are the level $\ell$ {\em dominant integral weights} (DIW)
of $\widehat{sl(n)}$.  Explicitly, we state

\begin{de}[State space]

For $\Lambda_i, i=0,\ldots,n-1$ the fundamental weights of
$\widehat{sl(n)}$, and some natural number $\ell$,
$$P(n;\ell)=\{ \Lambda=\sum_{i=0}^{n-1} m_i \Lambda_i |
m_i \in \zz, \sum_{i=0}^{n-1} m_i = \ell \}$$
is the set of integral weights,
and the state space is
$$P^+(n;\ell)=\{ \Lambda=\sum_{i=0}^{n-1} m_i \Lambda_i \in P(n;\ell) |
m_i \geq 0 \}.$$

\end{de}

Let $\epsilon_j=(0,\cdots,0,1,0,\cdots,0)$ be an $n$-dimensional
vector with $1$ in the $j^{\scriptstyle\mbox{th}}$ position,
$j=0,\ldots,n-1$, and
define the set ${\cal A}_{\ell}^+$:
$${\cal A}_{\ell}^+ = \{\nu=\sum_{i=0}^{n-1}\nu_i \epsilon_i |
\nu_i \in \zz, \nu_i \geq 0, \sum_{i=0}^{n-1} \nu_i = \ell \}.$$
The subscript $i$ of $\Lambda_i, \epsilon_i$ may be extended to
$i \in \zz$ by setting $\Lambda_i=\Lambda_j$ and
$\epsilon_i=\epsilon_j$ for $i\equiv j$ (mod $n$).

The configurations of states that label the eigenvectors of the
corner transfer matrix (CTM) are those that dominate the partition
function in the low temperature limit.  These low-temperature
configurations are best described by a set of {\em paths}
which we define below.

\begin{de}[Path, $\Lambda$-path]

A path is a pair $(\mu, \eta)$ such that
\begin{enumerate}
\item $\mu=(\mu_k)_{k\geq 0}$, $\mu_k \in P(n;\ell)$
\item $\eta=(\eta_k)_{k\geq 0}$, $\eta_k \in {\cal A}_{\ell}^+$
and
\item $\mu_{k+1}-\mu_k=\bar{\eta}_k \ \forall k$, where
$-:{\cal A}_{\ell}^+\rightarrow P_0$ is a $\zz$-linear map which maps
$\epsilon_j \mapsto \bar{\epsilon}_j=
\Lambda_{j+1}-\Lambda_j$.
\end{enumerate}
A $\Lambda$-path is a path as above, such that $\mu_k=\sigma^k (\Lambda)$
for $k>>0$ and where $\sigma(\Lambda_j)=\Lambda_{j+1} \, \forall j$.
\end{de}

\noindent In fact, the class of paths we shall focus on in this
paper are a subset of those described above.

\begin{de}[Restricted paths \cite{JMMO}]

Let $\Lambda, \Lambda'$ be dominant integral weights of level $l, l'$
respectively.
A $(\Lambda',\Lambda)$ path is defined as a pair of sequences
$(\mu,\eta)$ such that
\begin{enumerate}
\item  $(\mu-\Lambda',\eta)$ is a $\Lambda$-path, and
\item $\mu_k-$\v{$\eta_k$} $\in P^+(n;\ell')\, \forall k\geq 0$,
where \v{}$:{\cal A}\rightarrow P$ is a $\zz$-linear map which maps
$\epsilon_j \mapsto$ \v{$\epsilon_j$}=$\Lambda_j$.
\end{enumerate}
\end{de}

\noindent Unpacking the definitions a little, we note that
whereas a path is defined as a sequence of integral weights of
$\widehat{sl_n}$, the restricted paths are sequences of {\em dominant}
integral weights (often abbreviated as DIW).  This is the key
feature that distinguishes vertex from RSOS models.  We
can give an equivalent definition as follows of restricted paths
emphasizing this difference.

\begin{de}[Restricted paths, again]
For DIWs $\Lambda\in P^+_l, \,\Lambda'\in P^+_{l'}$, $(\mu,\eta)$
is a $(\Lambda',\Lambda)$-path iff
\begin{enumerate}
\item $\mu=(\mu_k)_{k\geq 0} \, \, \mu_k\in P_{l+l'}^+$,
\item $\mu_k=\Lambda'+\sigma^k(\Lambda)$ for $k>>0$,
\item $\eta=(\eta_k)_{k\geq 0} \, \, \eta_k\in {\cal A}_l^+$ and
\item $\mu_{k+1}-\mu_k=\bar{\eta}_k$,
\end{enumerate}
where $\bar{\cal A}_l$ and $\sigma(\Lambda)$ are defined above.
\end{de}

\def\egs(#1){\eta^{\scriptscriptstyle GS}_{#1}}

In this paper, we shall only consider the cases $l'=\ell-1, l=1$.
Therefore, only the specification of the sequence
$\eta_k\in\{\epsilon_j\}_{\scriptstyle 0\leq j\leq (n-1)}$ is
sufficient to characterize a $\Lambda$-path.  Let us denote the
set of all such restricted paths as ${\cal P}_\ell (\Lambda',\Lambda)$.

A {\em ground state path} for this class of (fundamental)
models we shall consider in regime III of the parameter space,
is a $(\Lambda',\Lambda)$
path with $\egs()=(\eta_{i+kn}=\epsilon_i)$ for $k \geq 0$ and
$0\leq i \leq(n-1)$.
In other words, $\mu_k=(\Lambda'-\Lambda_0)+\sigma^k(\Lambda_0)$
is a ground state path for any choice of $\Lambda'\in P^+(n;\ell)$.
The set of all paths in ${\cal P}_\ell(\Lambda',\Lambda)$ for which
$(\mu_k,\eta_k)_{k\geq L+1}$ is a ground state path, is called
a {\em finitized path}, and is denoted
${\cal P}_\ell^L(\Lambda',\Lambda)$.  It is customary to strip off
the semi-infinite tail of the sequence which is indistinguishable
from the ground state sequence.

In this paper, we shall describe a state
by a Young diagram, $\lambda=(\lambda_1,\ldots,\lambda_n)$, where
$\lambda_1\geq\cdots\geq\lambda_n$ and $\lambda_1-\lambda_n\leq \ell$.
Let the set of all such diagrams be denoted ${\cal Y}_{n,\ell}$.
A  dominant integral weight is represented by such a diagram, and
the map  wt:${\cal Y}_{n,\ell}\rightarrow P^+(n;\ell)$, defined as
$$\mbox{wt}(\lambda)=\sum_{i=0}^{n-1}
(\lambda_i-\lambda_{i+1})\Lambda_i,$$
where $\lambda_0:=\ell-(\lambda_1-\lambda_n)$.

\begin{figure}
\setlength{\unitlength}{0.00675in}
\begin{picture}(670,355)(0,-10)
\drawline(630,40)(650,40)(650,20)(630,20)
\drawline(630,20)(630,40)(610,40)
	(610,20)(630,20)
\drawline(650,40)(670,40)(670,20)(650,20)
\drawline(630,20)(630,0)(610,0)(610,20)
\drawline(650,20)(650,0)(630,0)
\drawline(670,20)(670,0)(650,0)
\drawline(330,145)(350,145)(350,125)(330,125)
\drawline(330,125)(330,145)(310,145)
	(310,125)(330,125)
\drawline(330,125)(330,105)(310,105)(310,125)
\drawline(430,240)(450,240)(450,220)(430,220)
\drawline(430,220)(430,240)(410,240)
	(410,220)(430,220)
\drawline(450,240)(470,240)(470,220)(450,220)
\drawline(430,220)(430,200)(410,200)(410,220)
\drawline(440,40)(460,40)(460,20)(440,20)
\drawline(440,20)(440,40)(420,40)
	(420,20)(440,20)
\drawline(440,20)(440,0)(420,0)(420,20)
\drawline(440,0)(460,0)(460,20)
\drawline(230,20)(230,40)(210,40)
	(210,20)(230,20)
\drawline(230,20)(230,0)(210,0)(210,20)
\drawline(525,150)(545,150)(545,130)(525,130)
\drawline(525,130)(525,150)(505,150)
	(505,130)(525,130)
\drawline(545,150)(565,150)(565,130)(545,130)
\drawline(525,130)(525,110)(505,110)(505,130)
\drawline(545,130)(545,110)(525,110)
\drawline(220,235)(240,235)(240,215)(220,215)
\drawline(220,215)(220,235)(200,235)
	(200,215)(220,215)
\drawline(310,340)(330,340)(330,320)(310,320)
\drawline(310,320)(310,340)(290,340)
	(290,320)(310,320)
\drawline(330,340)(350,340)(350,320)(330,320)
\drawline(245,205)(300,155)
\drawline(292.735,158.902)(300.000,155.000)(295.426,161.861)
\drawline(351,106)(406,56)
\drawline(398.735,59.902)(406.000,56.000)(401.426,62.861)
\drawline(441,210)(496,160)
\drawline(488.735,163.902)(496.000,160.000)(491.426,166.861)
\drawline(560,105)(615,55)
\drawline(607.735,58.902)(615.000,55.000)(610.426,61.861)
\drawline(161,100)(216,50)
\drawline(208.735,53.902)(216.000,50.000)(211.426,56.861)
\drawline(235,50)(290,105)
\drawline(285.757,97.929)(290.000,105.000)(282.929,100.757)
\drawline(140,150)(195,205)
\drawline(190.757,197.929)(195.000,205.000)(187.929,200.757)
\drawline(240,250)(295,305)
\drawline(290.757,297.929)(295.000,305.000)(287.929,300.757)
\drawline(340,155)(395,210)
\drawline(390.757,202.929)(395.000,210.000)(387.929,205.757)
\drawline(365,220)(280,220)
\drawline(288.000,222.000)(280.000,220.000)(288.000,218.000)
\drawline(580,20)(495,20)
\drawline(503.000,22.000)(495.000,20.000)(503.000,18.000)
\drawline(370,20)(285,20)
\drawline(293.000,22.000)(285.000,20.000)(293.000,18.000)
\drawline(170,20)(85,20)
\drawline(93.000,22.000)(85.000,20.000)(93.000,18.000)
\drawline(445,55)(500,110)
\drawline(495.757,102.929)(500.000,110.000)(492.929,105.757)
\drawline(475,130)(390,130)
\drawline(398.000,132.000)(390.000,130.000)(398.000,128.000)
\drawline(345,300)(400,250)
\drawline(392.735,253.902)(400.000,250.000)(395.426,256.861)
\drawline(130,115)(130,135)(110,135)
	(110,115)(130,115)
\drawline(40,45)(95,100)
\drawline(90.757,92.929)(95.000,100.000)(87.929,95.757)
\drawline(260,130)(175,130)
\drawline(183.000,132.000)(175.000,130.000)(183.000,128.000)
\put(0,20){\makebox(0,0)[lb]{\raisebox{0pt}[0pt][0pt]{\shortstack[l]{{
$ \, \, \emptyset$}}}}}
\end{picture}
\caption{\sf The oriented graph ${\cal G}$ with the Young diagrams at
the nodes representing DIWs.}
\end{figure}

We shall translate the definition of restricted paths
into the language of Young diagrams as follows.
Let us introduce an oriented graph ${\cal G}$, the set of whose nodes
is $P^+(n;\ell)$ and whose oriented bonds (or arrows) are drawn as
follows.  For $a,b\in P^+(n;\ell)$, we draw an arrow between $a$ and
$b$ if
$$b-a  = \bar{\epsilon}_i = \Lambda_{i+1}-\Lambda_i \in
\bar{\cal A}_1^+, \, \mbox{ for some }
i=0,\ldots,n-1.$$
$\bar{\epsilon}_j$, for $j=1,\ldots,n-1$ may be identified with the
weights of the fundamental representation of $sl_n(\cc)$.
In terms of Young diagrams, $\alpha,\beta\in {\cal Y}_{n,\ell}$,
with wt($\alpha$)$=a$ and wt($\beta$)$=b$ are connected by an arrow
from $\alpha$ to $\beta$ if $\beta$ is obtained by adding a box to
the $i+1^{\scriptstyle th}$ row of $\alpha$.  Thus, for some $k$,
$\mu_k=a=wt(\alpha)$, $\mu_{k+1}=b=wt(\beta)$ and $\bar{\epsilon}_i$
is described by the extra box in the $i+1^{\scriptstyle th}$ row.

\subsection{Paths as tableaux}

A {\em standard} Young tableau $T$ is a Young diagram
whose cells are occupied by integers that increase along rows and
columns.

\def\ho{\hat{0}}
\def\hi{\hat{1}}

\begin{de}[Lattice permutation]

A word or multiset permutation $w$ of the letters $0,\ldots,(n-1)$,
is called a {\em lattice permutation}
 if, on reading from left to
right, the number of occurences of
$i$ in $w$ is no less than that of $j$,
for all $i<j$.
\end{de}

For a given sequence of integers labelling a path, $\eta(p)$ is
a multiset permutation.  We shall encode such a path
$\eta(p)=(\eta_1,\ldots,\eta_L)$ as an array by the
following procedure.  We place
the integer $j$ in the $(i+1)^{\scriptstyle th}$ row of an array
iff $\eta_j=\hat{i}$.  For restricted paths which are sequences of DIWs,
the array thus formed is of partition shape, and are Young tableaux,
and are thus lattice permutations.  We can stick to normal
standard tableaux in this paper (not skew shapes) unless we
specify otherwise, with the first integer in $\eta(p)=\ho$.
To a path $p$, we denote by $T_p$ the tableau associated to it
by this procedure,and to any tableau $T$,
We call $w_T$ the lattice permutation corresponding to $T$,
so that $p=w_{T_p}$.

\medskip

\def\step(#1){\stackrel{\bar{\epsilon}_{#1}}{\rightarrow}}
\noindent {\bf Example.} The following sequence of Young diagrams

\medskip

\setlength{\unitlength}{0.00625in}
\begin{picture}(723,55)(0,-10)
\drawline(240,40)(260,40)(260,20)(240,20)
\drawline(240,20)(240,40)(220,40)
	(220,20)(240,20)
\drawline(240,20)(240,0)(220,0)(220,20)
\drawline(385,30)(405,30)(405,10)(385,10)
\drawline(385,10)(385,30)(365,30)
	(365,10)(385,10)
\drawline(470,40)(490,40)(490,20)(470,20)
\drawline(470,20)(470,40)(450,40)
	(450,20)(470,20)
\drawline(470,20)(470,0)(450,0)(450,20)
\drawline(555,40)(575,40)(575,20)(555,20)
\drawline(555,20)(555,40)(535,40)
	(535,20)(555,20)
\drawline(575,40)(595,40)(595,20)(575,20)
\drawline(555,20)(555,0)(535,0)(535,20)
\drawline(155,30)(175,30)(175,10)(155,10)
\drawline(155,10)(155,30)(135,30)
	(135,10)(155,10)
\drawline(665,40)(685,40)(685,20)(665,20)
\drawline(665,20)(665,40)(645,40)
	(645,20)(665,20)
\drawline(685,40)(705,40)(705,20)(685,20)
\drawline(665,20)(665,0)(645,0)(645,20)
\drawline(685,20)(685,0)(665,0)
\drawline(90,10)(90,30)(70,30)
	(70,10)(90,10)
\drawline(325,10)(325,30)(305,30)
	(305,10)(325,10)
\put(0,15){\makebox(0,0)[lb]{\raisebox{0pt}[0pt][0pt]{\shortstack[l]{{$
\emptyset$}}}}}
\put(50,15){\makebox(0,0)[lb]{\raisebox{0pt}[0pt][0pt]{\shortstack[l]{{
$\hspace{-.5cm}\step(0)$}}}}}
\put(110,15){\makebox(0,0)[lb]{\raisebox{0pt}[0pt][0pt]{\shortstack[l]{{
$\hspace{-.35cm}\step(0)$}}}}}
\put(190,15){\makebox(0,0)[lb]{\raisebox{0pt}[0pt][0pt]{\shortstack[l]{{
$\hspace{-.25cm}\step(1)$}}}}}
\put(275,15){\makebox(0,0)[lb]{\raisebox{0pt}[0pt][0pt]{\shortstack[l]{{
$\hspace{-.25cm}\step(2)$}}}}}
\put(340,15){\makebox(0,0)[lb]{\raisebox{0pt}[0pt][0pt]{\shortstack[l]{{
$\hspace{-.25cm}\step(0)$}}}}}
\put(425,15){\makebox(0,0)[lb]{\raisebox{0pt}[0pt][0pt]{\shortstack[l]{{
$\hspace{-.35cm}\step(1)$}}}}}
\put(505,15){\makebox(0,0)[lb]{\raisebox{0pt}[0pt][0pt]{\shortstack[l]{{
$\hspace{-.25cm}\step(0)$}}}}}
\put(620,15){\makebox(0,0)[lb]{\raisebox{0pt}[0pt][0pt]{\shortstack[l]{{
$\hspace{-.25cm}\step(1)$}}}}}
\put(720,15){\makebox(0,0)[lb]{\raisebox{0pt}[0pt][0pt]{\shortstack[l]{{
$\hspace{-.2cm}\step(1)$}}}}}
\end{picture}

\medskip
\noindent encodes the restricted path or sequence of DIWs
$$\begin{array}{llll}
\ell\Lambda_0  & \step(0)(\ell-1)\Lambda_0+ \Lambda_1
	       & \step(0)(\ell-2)\Lambda_0+2\Lambda_1
	       & \step(1)(\ell-2)\Lambda_0+ \Lambda_1+ \Lambda_2 \\
	       & \step(2)(\ell-1)\Lambda_0+ \Lambda_1
	       & \step(0)(\ell-2)\Lambda_0+2\Lambda_1
	       & \step(1)(\ell-2)\Lambda_0+ \Lambda_1+ \Lambda_2 \\
	       & \step(0)(\ell-3)\Lambda_0+2\Lambda_1+ \Lambda_2
	       & \step(1)\cdots &
\end{array}$$
as does the following Young tableau
$$\begin{array}{cccc}
1  &  2  &  5  &  7 \\
3  &  6  &  8  &    \\
4  &     &     &
\end{array}.$$

We also note that any tableau $T$ representing some path
$p\in{\cal P}_\ell^L$ contains $L$ boxes, and any {\em  partial
tableau} $T^{(k)}$ of $T$ is the sub-diagram that contains the first
$k$ integer entries.  For example, for $T$ as above, $L=8$ and
the partial tableau $T^{(5)}$ is
$$T^{(5)}=
\begin{array}{ccc}
1 & 2 & 5 \\
3 &   &   \\
4 &   &
\end{array}.$$

Also, for most of the paper we shall be focusing on the {\em vacuum
module} for the cosets ${(A^{(1)}_{n-1})_{\ell -1}
\otimes (A^{(1)}_{n-1})_{1}}
\over (A^{(1)}_{n-1})_{\ell}$ and therefore \cite{JMMO}, we
focus on ${\cal P}_\ell^L(\ell \Lambda_0,\Lambda_0)$.
In terms of Young diagrams this means that
a path begins at the empty diagram $\emptyset$ and after $L=kn$ steps
(for some positive integer $k$), comes back to $\emptyset$.
Therefore, $\eta$ contains an equal number of $\epsilon_j$s for
$j=0,1,\ldots,n-1$, and the shapes of the corresponding tableaux
are thus rectangular, $(k^n)$.

The physical model is defined by the Boltzmann weights, and these
are non-zero for a given specification of states on the
adjacent sites of the physical (two-dimensional) lattice.  This
condition on nearest neighbour states can be expressed in terms
of the oriented graph structure on $P^+(n;\ell)$.  Namely, for
some given natural number $f$, a(n ordered) pair of states can
occupy nearest neighbour sites on the lattice, if it takes at most
$f$ steps to go from $a$ to $b$ in ${\cal P}$.  $f$ is called the degree
of fusion, and as mentioned before, we shall only consider the case
$f=1$ in this paper.

\section{Functionals on paths and tableau}

We define {\em an energy function} $H$ on a path as follows.
\begin{de}[energy of a path]
The energy of a path $p$, $E(p)$ is given by the the evaluation of
 a functional
$H(p)$,
\begin{equation}
H(p)=\sum_{j=1}^L j \theta(\eta_{j}-\eta_{j -1}),
\label{energy-E}
\end{equation}
relative to its evaluation on the ground state path $\bar{p}$:
\begin{equation}
E(p)=H(p)-H(\bar{p})
\end{equation}
and $\theta$ is the step function given by
\begin{equation}
\theta(z) = \left\{
\begin{array}{ll}
0 \qquad & (z  >   0) \\
1        & (z \leq 0).
\end{array}
\right.
\label{step}
\end{equation}
It is also useful to state the following convention for later
usage: for a section of a path $(\mu_i,\mu_{i+1}\mu_{i+2})$ given by
the three integral weights $(\lambda_1,\Lambda_2=\Lambda_1+\bar{\epsilon}_j,
\Lambda_3=\Lambda_1+\Lambda_2+\bar{\epsilon}_k)$,
one can define a local energy function:
$$H(\Lambda_1,\Lambda_1+\bar{\epsilon}_j,\Lambda_1+\bar{\epsilon}_j+
\bar{\epsilon}_k):=\theta(k-j).$$

\end{de}


Note that for a given $p$, the contribution to $E(p)$ of each step
$j$ of the path
is {\em zero} if the integer $(j+1)$ lies {\em below} $j$ in $T_p$.
With this property in mind, we introduce
two integer-valued functionals or {\em statistics} on tableaux,
$c(T)$ and ${\bf p}(T)$.

\def\des{{\cal D}}

\begin{de}[Charge of a tableau \cite{L-Sch}]
For a row word $\pi_T=a_1a_2\ldots a_L$ of a tableau $T$, assign an
{\em index} $i$ to each $a_i$ as follows.
$$\mbox{index}(a_j=x+1):=\left \{
\begin{array}{lll}
0                  & \mbox{if} & x=0 \\
\mbox{index}(x)+1  & \mbox{if} & a_i=x \, \mbox{and} \, i<j \\
\mbox{index}(x)    & \mbox{otherwise} &.
\end{array}              \right.
$$
The charge $c(\pi_T)$ of a word $w=\pi_T$ is
$$\mbox{charge}(\pi_T):=\sum_{i=1}^L \mbox{index}(a_i),$$
and the charge of a tableau $c(T)=c(\pi_T)$.
\end{de}

\noindent In words, if the number $x$ has index $i$, then $x+1$ will have
index $i+1$ if it is to the {\em right} of $x$, and $i$ otherwise.

In order to define the statistic ${\bf p}(T)$ it is necessary to
set up the following definitions.
For any permutation $w$ of $\{1,2,\ldots,L\}$ introduce the following
sets:
\begin{de}[$\des( {w} ),{\des}^t({T})$] \hfill

\noindent (i) If $j$ and $j+1$ are both in $w$,
$$\des( {w} )=\{ j | j+1 \, \mbox{lies to the {\em left} of}\, j \,
 \mbox{in} \, w\}.$$
We also set $d(w)=|\des({w})|$ and $des(w)=\sum_{j\in\des({w})}j$.

\noindent (ii) For a tableau $T$,
$$\des^t(T)=\{j|j+1 \, \mbox{lies {\em below} } \, j\, \mbox{in}
\, T\}.$$
\end{de}

\begin{de}[Row word of a tableau \cite{greene}]

For a tableau $T$, the row word $\pi_T$ of $T$ is the permutation
$$\pi_T=R_n R_{n-1} \ldots R_1$$
where $R_i, i=1,\ldots,n$ are the rows of $T$ read from left to right.

\end{de}

\noindent {\bf Example.}
$$\begin{array}{cc}
{T=\begin{array}{cccc}
	1 & 3 & 5 & 6 \\
	2 & 4 & 7 &   \\
	8 &   &   &   \\
\end{array}}& {\pi_T=82471356}
\end{array}$$

\noindent Note that $\des^t(T)=\des(\pi_T)$, and
we shall drop the superscript on $\des^t$ and use the same
notation $\des$ to refer to the map on both domains,
words and tableaux.  Also, set $d(T)=d(\pi_T)$ and $des(T)=des(\pi_T)$.

\noindent {\bf Example.}
For $T$ as above, $\des(T) = \des(\pi_T) = \{1,3,6,7\}$, $d(\pi_T)=4$
and $des(\pi_T)=17$.

\noindent {\bf Remark.} The row word of a tableau is a convenient index
\cite{greene} for a Knuth (or plactic) equivalence class.
This is a set of words that
are related to each other by the Knuth relations \cite{knuth}
(or plactic monoid).
Every word in each Knuth class gives the same $P$ tableau under the
Robinson-Schensted-Knuth correspondence.

\begin{de}[statistic {\bf p}(T) \cite{thomas}]
For a tableau $T$, introduce the set
$${\cal E}_T=\{j | j+1 \mbox{lies to the right of }\, j\,
\mbox{ in }\, T\},$$
and define ${\bf p}(T):=\sum_{j\in {\cal E}_T}j$.
\end{de}

\noindent {\bf Remark.} Observe that for a standard tableau $T$,
$\des(T)\cap{\cal E}_T=\emptyset$.  In words, for every entry $j$
in a standard tableau, $(j+1)$ lies either to the right of or below
$j$, but not both.

\noindent {\bf Example.}
For $T$ as above, the indices for $1,2,\ldots,8$ are $0,0,1,1,2,3,3,3$
respectively, and $c(T)=13$.

\noindent {\bf Remark.}  The statistics $c(T)$ and {\bf p}$(T)$ are
invariant under the Knuth relations.

\subsection{Energy of a path as statistic {\bf p}}

Observe that for a path
$\egs():=(0,1,\ldots,n-1,0,1,\ldots,n-1,\ldots,0,1,\ldots, n-1)$, of
length $L=nx$ (say), if the corresponding tableau is $T_{\scriptstyle gs}$,
then ${\cal E}_{T_{\scriptstyle gs}}=\{n,2n,\ldots,(x-1)n\}$.
Note, that if we treat every path $p$ as a {\em finite} one, then
$L\not\in {\cal E}_{T_p}$.

\begin{pr}
For a path $\eta=(\eta_1,\ldots,\eta_L)$ encoded as a Young tableau
$T$ with $|\mbox{sh}T|=L=nx$ for $x$ a natural number.
  The energy of the path $E(\eta)$ is then
$$E(\eta)={\bf p}(T)-{L\over 2}({L\over n}-1).$$
\end{pr}
\noindent {\em Proof:} Observe ${\bf p}(T_p)=H(p)$. {\hfill $\Box$}

For ease of computation, it is convenient to note

\begin{pr} \hfill

\noindent (i) c(T)$={1\over 2}L(L-1) - L d(T) + des(T)$;

\noindent (ii) {\bf p}(T)$={1\over 2}L(L-1) - des(T)$.

\end{pr}

\noindent {\em Proof:} (i) Observe that the contributions to $c$ come
from the sum $0+1+\cdots +\bigl(L-d(\pi_T)-1\bigr)$ and also
from the set $\des(\pi_T)$.  Namely,
if $x$ is the $i^{\scriptstyle th}$ element of $\des(\pi_T)$, then
index($x$)$=(x-i)$.  Thus
$$c=0+1+\cdots+(L-d-1)+\bigl( des(\pi_T) - (1+2+\cdots+d)\bigr),$$
and the result follows.

\noindent (ii) See the remark following the definition of {\bf p}(T).

\hfill{$\Box$}

For future reference we shall introduce the following notation.

\begin{de}
\label{i-tab}
For a standard tableau $T$, with $|sh \, T|=L$,
$$\mbox{{\sf I}}(T,L):=\bigl(des(T)-{1\over 2}L d(T)\bigr).$$
\end{de}

\noindent Therefore, we can write
\begin{equation}
\label{cp-defs}
\begin{array}{lll}
c(T)&=&{1\over 2}L\bigl(L-1-d(T)\bigr)+\mbox{\sf I}(T,L)\\
{\bf p}(T)&=&{1\over 2}L\bigl(L-1-d(T)\bigr)-\mbox{\sf I}(T,L)
\end{array}.
\end{equation}

\subsection{An involution on tableaux: evacuation}

The involution we shall describe is a special case of a family of operations
which are indexed by points $p\in Z\times Z$ occupied by an ordered set of
objects on which they act injectively.  These were defined by Sch\"utzenberger
and are called {\em jeu de taquin}.  For the involution $S$ on standard
tableaux, (also called {\em evacuation}), the point in question is the
top left corner $(1,1)$, {\em e.g.}, the cell
occupied by the number $1$ in our example above.

The operation $S$ is defined by the following set of moves.  First, remove
the entry in cell $(1,1)$.  Move the smaller of the two integers immediately
below or to the right of it to occupy the cell $(1,1)$.  This leaves a
vacancy in the cell from which the integer was moved.  Then look for the
smallest integer immediately below or to the right of this freshly vacated
cell, and move that one in to squat in it.  This leaves a new cell empty.
In general, whenever the cell $(i,j)$ is empty, and $t(i_1,j_1)=
\mbox{min}(t(i+1,j),t(i,j+1))$, slide the integer from $(i_1,j_1)$ to
$(i,j)$.  And so on, until there are no more cells to the right and below
the cell vacated.  In this last cell insert the number $n$ with parentheses
around it, to distinguish it from the $n$ that is already present in the
tableau.  Now, remove the (new) integer from the cell $(1,1)$, and repeat
this game, inserting $(n-1)$ in the last cell.  And
so on, until all the entries of the original tableau are deleted, and we have
parentheses around the integers in the new standard tableau, which we
then remove.

\noindent {\bf Example. [evacuation]}

In the tableau $T$ above, remove the entry from the cell $(1,1)$ and
start the game:
\[
\begin{array}{cccc}
\cdot & 2 & 4 & 8 \\
3 & 6 & 9 &   \\
5 & 7 &   &
\end{array}
\rightarrow
\begin{array}{cccc}
2 & \cdot & 4 & 8 \\
3 & 6 & 9 &   \\
5 & 7 &   &
\end{array}
\rightarrow
\begin{array}{cccc}
2 & 4 & \cdot & 8 \\
3 & 6 & 9 &   \\
5 & 7 &   &
\end{array}
\rightarrow
\begin{array}{cccc}
2 & 4 & 8 & \bullet\\
3 & 6 & 9 &   \\
5 & 7 &   &
\end{array}.
\]
Replace the $\bullet$ by $(9)$.  Iterating this one more time, we get
\[
\begin{array}{cccc}
\cdot & 4 & 8 & (9) \\
3 & 6 & 9 &   \\
5 & 7 &   &
\end{array}
\rightarrow
\begin{array}{cccc}
3 & 4 & 8 & (9) \\
\cdot & 6 & 9 &   \\
5 & 7 &   &
\end{array}
\rightarrow
\begin{array}{cccc}
3 & 4 & 8 & (9) \\
5 & 6 & 9 &   \\
\cdot & 7 &   &
\end{array}
\rightarrow
\begin{array}{cccc}
3 & 4 & 8 & (9) \\
5 & 6 & 9 &   \\
7 & \bullet &   &
\end{array}.
\]
After a couple of more steps, we have

\[
\begin{array}{cccc}
3 & 4 & 8 & (9) \\
5 & 6 & 9 &   \\
7 & (8) &   &
\end{array}
\rightarrow
\begin{array}{cccc}
4 &  6 &  8 & (9)\\
5 &  9 & (7) &   \\
7 & (8) &   &
\end{array}
\rightarrow
\begin{array}{cccc}
5  & 6 & 8 & (9)\\
7  & 9 & (7) &   \\
(6) & (8) &   &
\end{array},
\]
and so on until
\[
T^S=
\begin{array}{cccc}
1 & 3 & 5 & 9 \\
2 & 4 & 7 &   \\
6 & 8 &   &
\end{array}.
\]
\noindent  Check that $S$ is an involution, {\em i.e.} $S^2=1$,
or $(T^S)^S=T$.

\begin{pr}
For tableaux $T$ and $T^S$ related by the evacuation involution,
$c(T)={\bf p}(T^S)$.
\end{pr}

\noindent {\em Proof}: See \cite{Butler}. {\large Give proof here}.
{\hfill $\Box$}

\noindent {\bf Remark.}  As a trivial corollary to the above, note
that
\begin{equation}
\label{I-inv}
\mbox{\sf I}(T,L)=-\mbox{\sf I}(T^S,L).
\end{equation}

\section{Introducing Rigged Configurations.}

Let $\lambda$ and $\mu$ be partitions such that
$|\lambda|=|\mu|$ and the number of parts $\ell(\lambda)$ of
$\lambda$ is no bigger than $n$.  Since we are interested in
the cosets $(\widehat{sl(n)})_{\ell-f}\otimes
(\widehat{sl(n)})_{f}/(\widehat{sl(n)})_{\ell}$ for
the fusion level $f=1$ in this paper, we choose $\mu=(1^L)$ to
indicate that the state space is chosen from the summands in the
decomposition of an $L$-fold tensor product of fundamental
representations of $sl_n(\mbox{\bf C})$.

For a given $\lambda$ one can define a sequence of partitions
$$\vec{\nu}=(\nu^{(0)},\nu^{(1)},\ldots,\nu^{(n-1)})$$
which satisfy the condition
\begin{equation}
\label{sum-rule}
|\nu^{(a)}|=\sum_{k\geq a+1}\lambda_k.
\end{equation}
We set $\nu^{(0)}=\mu=1^L$.
We shall call the rows of a partition $\nu^{(a)}$ {\em strings of
colour} $a$.  Let $l_*$ be the maximum allowed length of string,
$s_j(\nu^{(a)})$ be the number of strings of colour $a$ and
length $j$, and $Q_j^{(a)}(\nu)$ be the number of cells in the
first $j$ columns of diagram $\nu^{(a)}$, {\em i.e.}
$Q_j^{(a)}=\sum_{i=1}^{l_*}$min$(i,j)s_i^{(a)}$.

To every string in $\nu^{(a)}$ is assigned an integer
$J_{j,\alpha}^{(a)}$, $\alpha=1,\ldots,s_j(\nu^{(a)}), a=1,\ldots,n-1$
and $j=1,\ldots,l_*$, called a {\em rigging}.  These integers
$J_{j,\alpha}^{(a)}$ are
chosen from the set $\{0,1,\ldots,h_j^{(a)}\}$.  The maximum
allowed integer, $h_j^{(a)}$ is called the number of {\em holes} of
colour $a$ and length $j$, and is determined by:
\begin{equation}
\label{holes}
h_j^{(a)}=Q_j^{(a-1)}-2 Q_j^{(a)}+Q_j^{(a+1)},
\end{equation}
We must have $h_j^{(a)}\geq 0$ for a configuration to be {\em admissible}.

\begin{de}[rigged configurations]

A configuration of partitions $\nu$ whose rows are
indexed by non-negative ({\em i.e., admissible}) integers
$J_{j,\mu}^{(a)}, \mu=1,\ldots,s_j^{(a)}$
no larger than $h_j^{(a)}$ for $1,\leq j\leq l_*$
where $h_j^{(a)}$ is calculated as above
is called a {\em rigged configuration}.  We shall
denote a rigged configuration by $(\vec{\nu},\{J\})$ where $\{J\}$
denotes the set of rigging integers.
\end{de}

The total number of admissible rigged configurations is given by
\begin{equation}
\sum_{\{\nu\}}\prod_{a=1}^{n-1}\prod_{j=1}^{l_*} \biggl(
\begin{array}{c}
s_j^{(a)}+h_j^{(a)}\\
s_j^{(a)}
\end{array}\biggr).
\end{equation}

Recall that the number of holes is calculated once the
number of
strings is known, using
$$h_j^{(a)}=Q_j^{(a-1)}-2Q_j^{(a)}+Q_j^{(a+1)}=L\delta_{a1}-
\sum_b G_{ab} Q_j^{(b)},$$
where  $G_{ab}$ are the matrix elements of the Cartan matrix
of
$sl_n$.  Inverting the above equation, we get
$$Q_j^{(a)}=L G^{-1}_{a1}-\sum_b G^{-1}_{ab} h_j^{(b)},$$
which, together with the relation
$Q_{j}^{(a)}-Q_{j-1}^{(a)}=\sum_{k\geq j} s_k^{(a)}$, may be
used
to express the number of strings in terms of the number of
holes.
Note,
\begin{equation}
\label{bar-h}
\bar{h}_j^{(a)}:=h_j^{(a)}-h_{j+1}^{(a)}=\sum_b \sum_{k\geq j+1}
G_{ab} s_k^{(b)},
\end{equation}
so that
\begin{equation}
\label{s-h}
\begin{array}{ccc}
s_j^{(a)}&=&(Q_j^{(a)}-Q_{j-1}^{(a)})-(Q_{j+1}^{(a)}-Q_j^{(a)}) \\
&=&LG^{-1}_{a1}\delta_{j1} - \sum_b \sum_k G^{-1}_{ab}C_{jk}h_k^{(b)},
\end{array}
\end{equation}
where $C_{jk}=(2\delta_{jk}-\delta_{j\,(k-1)}-\delta_{j\,(k+1)})$.

\noindent {\bf Remark.}
In the literature on counting string
solutions to Bethe equations, distinct integers
are assigned to strings \cite{Tak}, unlike the rigging
defined here.  The relation between
the distinct (also called fermionic) (half-)integer branches
$I_{j,\alpha}^{(a)}$ and the not necessarily distinct rigging
integers $J_{j,\alpha}^{(a)}$ is
given by
\begin{equation}
\label{takahashi}
I_{j,\alpha}^{(a)} = J_{j,\alpha}^{(a)} +
\alpha - {1\over2}(h_j^{(a)}+s_j^{(a)}+1).
\end{equation}

\section{The Kerov-Kirillov-Reshetikhin Algorithm}

In a remarkable paper \cite{KKR}, a bijection between standard Young
tableaux and rigged configurations was established.  In \cite{KR}
and \cite{K-JGP} this was extended to a charge preserving bijection
between tableau of shape $\lambda$ and weight $\mu$ a partition.  We shall
concentrate only on standard tableaux for the purpose of this paper.
We shall describe this bijection in this section.
Let SYT($\lambda$) be the set of standard Young tableau of shape
$\lambda$ and ${\cal R}_{\lambda}$ be the set of
admissible rigged configurations
$\{\vec{\nu},\{J\}\}$ satisfying the condition (\ref{sum-rule}).
We shall describe the monomorphisms (injective maps)
$$\tau:{\cal R}_{\lambda}\longrightarrow SYT(\lambda)\,\mbox{and}\,
\rho:SYT(\lambda)\longrightarrow {\cal R}_{\lambda},$$
with $\tau\circ\rho=\rho\circ\tau=$id.  We shall usually drop the
subscript on ${\cal R}$.

\subsection{$\rho:SYT(\lambda)\longrightarrow {\cal R}.$}

We shall describe the map from $w_T$, $T\in SYT(\lambda)$ to an
admissible rigged configuration.  The construction is inductive --
let there be a rigged configuration constructed by the first $k-1$ steps
of the sequence, {\em i.e.} $a_1 a_2 \ldots a_{k-1}$.  We shall describe
how to make the $k^{\scriptstyle th}$ step.  For the description of
the map $\rho$, it is convenient to
arrange the integers in the rows of the same length $j$
in $\nu^{(a)} \, \forall j$ such that the integer riggings read from
the top are in {\em weakly decreasing} order.

\begin{de}[Edge string]
A row of $\nu^{(a)}$ for any $a=1,\ldots,(n-1)$ is called
an {\em edge string} if its integer rigging is the maximum allowed,
{\i.e.} $J_{n,\alpha}^{(a)}=h_n^{(a)}$.
\end{de}

\noindent{\bf Remark.} In \cite{KKR,KR}, this is referred to as
a singular string, while in \cite{K-JGP}, it is called a special
string.

Let $a_k=a_*, 0\leq a_*\leq n-1$.  In the diagram $\nu^{(a_*)}$, consider
the {\em longest and highest edge string}, and let this
length be $\zeta^{a_*}$.  Note that $\zeta^{a_*}$ could be $0$.
For each of the diagrams $\nu^{(b)}$,
$0<b<a_*$, let $\zeta^b$ be the length of the longest and
highest edge string such that $\zeta^b \leq \zeta^{b+1}$.
Add a cell to the right of each of
these rows $\zeta^{b},j=1,\ldots a_*$, and calculate the values of
$h_x^{(a)}$ for
$x=1,\ldots,l_*$ and $a=1,\ldots,n-1$
for the new set of diagrams $\vec{\nu}$ after $k$ steps.  Note that
if $\zeta^b=0$, we create a string of length $1$.
Set the integer entries of the chosen elongated strings to be such
that they are again edge, {\em i.e.,}
$J_{\zeta^j +1,\alpha}^{(a)}=h_{\zeta^j +1}^{(a)}$, the maximum
allowed, for $a=1,\ldots a_*$.  Also make $\nu^{(0)}=(1^k)$.
Let the integers in all the other strings remain unchanged.
Rearrange the
resulting configuration, so that the integers are again in
non-increasing order.  This is the new rigged configuration
after $k$ steps.

It is easy to work out (see \cite{KR}),
that the change in $h_j^{(a)}$ because
of the change in the number of cells (and in $Q_j^{(a)}$)
in the sequence of diagrams
$\vec{\nu}$ is summed up as follows:
$$h_j^{(a)}\mapsto h_j^{(a)}+\left \{
\begin{array}{rc}
0 &(j\geq \zeta^{a+1})\\
-1&(\zeta^{a}\leq j < \zeta^{a+1})\\
+1&(\zeta^{a-1}\leq j < \zeta^{a})\\
0 &(j < \zeta^{a-1})
\end{array}
\right.
$$
This ensures that the new configuration thus constructed
is an admissible one, {\em i.e.} $h_j^{(a)}\geq 0$.

\noindent {\bf Example.} Consider the following rigged configuration,
formed after $7$ steps:

\medskip

\centerline{\setlength{\unitlength}{0.0075in}%
\begin{picture}(230,60)(35,720)
\thicklines
\put( 80,760){\line( 1, 0){ 40}}
\put(120,760){\line( 0, 1){ 20}}
\put(120,780){\line(-1, 0){ 40}}
\put( 80,780){\line( 0,-1){ 60}}
\put( 80,720){\line( 1, 0){ 20}}
\put(100,720){\line( 0, 1){ 60}}
\put( 80,740){\line( 1, 0){ 20}}
\put(180,760){\line( 0,-1){ 20}}
\put(180,740){\line(-1, 0){ 20}}
\put(160,740){\line( 0, 1){ 40}}
\put(160,780){\line( 1, 0){ 20}}
\put(180,780){\line( 0,-1){ 20}}
\put(180,760){\line(-1, 0){ 20}}
\put(240,780){\line( 0,-1){ 20}}
\put(240,760){\line( 1, 0){ 20}}
\put(260,760){\line( 0, 1){ 20}}
\put(260,780){\line(-1, 0){ 20}}
\put( 35,760){\makebox(0,0)[lb]{\raisebox{0pt}[0pt][0pt]{$(1^7)$}}}
\put( 85,765){\makebox(0,0)[lb]{\raisebox{0pt}[0pt][0pt]{\rm 0}}}
\put( 85,745){\makebox(0,0)[lb]{\raisebox{0pt}[0pt][0pt]{\rm 2}}}
\put( 85,725){\makebox(0,0)[lb]{\raisebox{0pt}[0pt][0pt]{\rm 1}}}
\put(105,735){\makebox(0,0)[lb]{\raisebox{0pt}[0pt][0pt]{\rm 3}}}
\put(125,765){\makebox(0,0)[lb]{\raisebox{0pt}[0pt][0pt]{\rm 1}}}
\put(165,765){\makebox(0,0)[lb]{\raisebox{0pt}[0pt][0pt]{\rm 0}}}
\put(165,745){\makebox(0,0)[lb]{\raisebox{0pt}[0pt][0pt]{\rm 0}}}
\put(185,755){\makebox(0,0)[lb]{\raisebox{0pt}[0pt][0pt]{\rm 0}}}
\put(245,765){\makebox(0,0)[lb]{\raisebox{0pt}[0pt][0pt]{\rm 0}}}
\put(265,765){\makebox(0,0)[lb]{\raisebox{0pt}[0pt][0pt]{\rm 0}}}
\end{picture}}

\medskip

Suppose the next step is a $2$.  This means we have to choose the
{\em longest} edge row in $\nu^{(2)}$(length $1$), and the longest
edge row in $\nu^{(1)}$ which is shorter or as long as the one we
just picked out
in $\nu^{(2)}$ (length $0$).    To each of
these, add a cell to the right, and insert the value of the maximum
allowed integer.  The configuration thus formed is:

\medskip

\centerline{\setlength{\unitlength}{0.0075in}%
\begin{picture}(235,80)(30,700)
\thicklines
\put( 80,760){\line( 1, 0){ 40}}
\put(120,760){\line( 0, 1){ 20}}
\put(120,780){\line(-1, 0){ 40}}
\put( 80,780){\line( 0,-1){ 60}}
\put( 80,720){\line( 1, 0){ 20}}
\put(100,720){\line( 0, 1){ 60}}
\put( 80,740){\line( 1, 0){ 20}}
\put(180,760){\line( 0,-1){ 20}}
\put(180,740){\line(-1, 0){ 20}}
\put(160,740){\line( 0, 1){ 40}}
\put(160,780){\line( 1, 0){ 20}}
\put(180,780){\line( 0,-1){ 20}}
\put(180,760){\line(-1, 0){ 20}}
\put(240,780){\line( 0,-1){ 20}}
\put(240,760){\line( 1, 0){ 20}}
\put(260,760){\line( 0, 1){ 20}}
\put(260,780){\line(-1, 0){ 20}}
\put( 80,720){\line( 0,-1){ 20}}
\put( 80,700){\line( 1, 0){ 20}}
\put(100,700){\line( 0, 1){ 20}}
\put(180,780){\line( 1, 0){ 20}}
\put(200,780){\line( 0,-1){ 20}}
\put(200,760){\line(-1, 0){ 20}}
\put( 85,765){\makebox(0,0)[lb]{\raisebox{0pt}[0pt][0pt]{\rm 0}}}
\put( 85,745){\makebox(0,0)[lb]{\raisebox{0pt}[0pt][0pt]{\rm 2}}}
\put( 85,725){\makebox(0,0)[lb]{\raisebox{0pt}[0pt][0pt]{\rm 2}}}
\put(125,765){\makebox(0,0)[lb]{\raisebox{0pt}[0pt][0pt]{\rm 1}}}
\put(165,765){\makebox(0,0)[lb]{\raisebox{0pt}[0pt][0pt]{\rm 0}}}
\put(165,745){\makebox(0,0)[lb]{\raisebox{0pt}[0pt][0pt]{\rm 0}}}
\put(245,765){\makebox(0,0)[lb]{\raisebox{0pt}[0pt][0pt]{\rm 0}}}
\put(265,765){\makebox(0,0)[lb]{\raisebox{0pt}[0pt][0pt]{\rm 0}}}
\put( 30,760){\makebox(0,0)[lb]{\raisebox{0pt}[0pt][0pt]{$(1^8)$}}}
\put( 85,705){\makebox(0,0)[lb]{\raisebox{0pt}[0pt][0pt]{\rm 1}}}
\put(105,730){\makebox(0,0)[lb]{\raisebox{0pt}[0pt][0pt]{\rm 2}}}
\put(185,745){\makebox(0,0)[lb]{\raisebox{0pt}[0pt][0pt]{\rm 1}}}
\put(205,765){\makebox(0,0)[lb]{\raisebox{0pt}[0pt][0pt]{\rm 0}}}
\end{picture}}

\medskip

\subsection{$\tau:{\cal R}\longrightarrow SYT(\lambda).$}

Let us now describe how to go from a rigged configuration to a
lattice permutation $w_T$, from which it is straightforward to recover
the appropriate Young tableau $T$.

For this map, we shall order the integers in the rows of equal length
so that they are {\em weakly increasing} from the top downwards.
The idea is to read off from a given rigged configuration
$(\vec{\nu},\{J\})$, a given number
$r$ called the {\em rank} of a configuration, and
give a prescription to modify the configuration, called
a {\em ramification rule}.  This
modification gives rise to a new configuration
$\widetilde{(\vec{\nu},\{J\})}$ with a new rank, $\tilde{r}$,
and so on.  This sequence
of ranks gives a lattice permutation.

In the diagram $\nu^{(1)}$,
consider the {\em shortest and lowest edge string}
of length $\xi^1\geq 1$.  Form a chain of length
$\xi^1\leq\xi^2\leq\ldots\xi^r$, such that there are no
edge strings of lengths greater than or equal to $\xi^r$ in the
diagram $\nu^{(r+1)}$. If $\nu^{(0)}=(1^j)$, set $a_j=r$.
Now delete the rightmost cell from each of these
chosen edge strings, and also
set $\nu^{(0)}=(1^{j-1})$.
Recompute the set of $h_x^{(a)}$ for the new configuration.  Note that
strings of length $\xi^i=1$ disappear.  Make the newly shortened
strings {\em edge}, {\em i.e.,} set the integers
$J_{\xi^i -1,\alpha}^{(a)}=h_{\xi^i -1}^{(a)}$,
the maximum allowed, for $i=1,\ldots,r$,
and leave the rest of the integer labels unchanged.

This gives rise to the new rigged configuration, and we may now
read off the value of the rank $a_{j-1}$.
We can thus construct the word $u=a_1, a_2,\ldots a_L$.
It is easy to check that under the changes
of configurations described above,
$$h_j^{(a)}\mapsto h_j^{(a)}+\left \{
\begin{array}{rc}
0 &(j\geq \xi^{a+1})\\
+1&(\xi^{a}\leq j < \xi^{a+1})\\
-1&(\xi^{a-1}\leq j < \xi^{a})\\
0 &(j < \xi^{a-1})
\end{array}
\right.
$$
so that $h_i^{(a)}$ remains non-negative.

\noindent {\bf Example.} Consider the following configuration:

\medskip

\centerline{\setlength{\unitlength}{0.0075in}%
\begin{picture}(235,80)(30,700)
\thicklines
\put( 80,760){\line( 1, 0){ 40}}
\put(120,760){\line( 0, 1){ 20}}
\put(120,780){\line(-1, 0){ 40}}
\put( 80,780){\line( 0,-1){ 60}}
\put( 80,720){\line( 1, 0){ 20}}
\put(100,720){\line( 0, 1){ 60}}
\put( 80,740){\line( 1, 0){ 20}}
\put(180,760){\line( 0,-1){ 20}}
\put(180,740){\line(-1, 0){ 20}}
\put(160,740){\line( 0, 1){ 40}}
\put(160,780){\line( 1, 0){ 20}}
\put(180,780){\line( 0,-1){ 20}}
\put(180,760){\line(-1, 0){ 20}}
\put(240,780){\line( 0,-1){ 20}}
\put(240,760){\line( 1, 0){ 20}}
\put(260,760){\line( 0, 1){ 20}}
\put(260,780){\line(-1, 0){ 20}}
\put( 80,720){\line( 0,-1){ 20}}
\put( 80,700){\line( 1, 0){ 20}}
\put(100,700){\line( 0, 1){ 20}}
\put(180,780){\line( 1, 0){ 20}}
\put(200,780){\line( 0,-1){ 20}}
\put(200,760){\line(-1, 0){ 20}}
\put( 85,765){\makebox(0,0)[lb]{\raisebox{0pt}[0pt][0pt]{\rm 0}}}
\put( 85,745){\makebox(0,0)[lb]{\raisebox{0pt}[0pt][0pt]{\rm 1}}}
\put( 85,725){\makebox(0,0)[lb]{\raisebox{0pt}[0pt][0pt]{\rm 2}}}
\put(125,765){\makebox(0,0)[lb]{\raisebox{0pt}[0pt][0pt]{\rm 1}}}
\put(165,765){\makebox(0,0)[lb]{\raisebox{0pt}[0pt][0pt]{\rm 0}}}
\put(245,765){\makebox(0,0)[lb]{\raisebox{0pt}[0pt][0pt]{\rm 0}}}
\put(265,765){\makebox(0,0)[lb]{\raisebox{0pt}[0pt][0pt]{\rm 0}}}
\put( 30,760){\makebox(0,0)[lb]{\raisebox{0pt}[0pt][0pt]{$(1^8)$}}}
\put( 85,705){\makebox(0,0)[lb]{\raisebox{0pt}[0pt][0pt]{\rm 2}}}
\put(105,730){\makebox(0,0)[lb]{\raisebox{0pt}[0pt][0pt]{\rm 2}}}
\put(185,745){\makebox(0,0)[lb]{\raisebox{0pt}[0pt][0pt]{\rm 1}}}
\put(205,765){\makebox(0,0)[lb]{\raisebox{0pt}[0pt][0pt]{\rm 0}}}
\put(165,745){\makebox(0,0)[lb]{\raisebox{0pt}[0pt][0pt]{\rm 1}}}
\end{picture}}

\medskip

\noindent Note that the rank of the above is $3$, and we delete cells
to obtain

\medskip

\centerline{\setlength{\unitlength}{0.0075in}%
\begin{picture}(175,60)(30,720)
\thicklines
\put( 80,760){\line( 1, 0){ 40}}
\put(120,760){\line( 0, 1){ 20}}
\put(120,780){\line(-1, 0){ 40}}
\put( 80,780){\line( 0,-1){ 60}}
\put( 80,720){\line( 1, 0){ 20}}
\put(100,720){\line( 0, 1){ 60}}
\put( 80,740){\line( 1, 0){ 20}}
\put(180,780){\line( 1, 0){ 20}}
\put(200,780){\line( 0,-1){ 20}}
\put(200,760){\line(-1, 0){ 20}}
\put(180,780){\line( 0,-1){ 20}}
\put(180,760){\line(-1, 0){ 20}}
\put(160,760){\line( 0, 1){ 20}}
\put(160,780){\line( 1, 0){ 20}}
\put( 85,765){\makebox(0,0)[lb]{\raisebox{0pt}[0pt][0pt]{\rm 0}}}
\put( 85,745){\makebox(0,0)[lb]{\raisebox{0pt}[0pt][0pt]{\rm 1}}}
\put( 85,725){\makebox(0,0)[lb]{\raisebox{0pt}[0pt][0pt]{\rm 2}}}
\put(125,765){\makebox(0,0)[lb]{\raisebox{0pt}[0pt][0pt]{\rm 1}}}
\put(165,765){\makebox(0,0)[lb]{\raisebox{0pt}[0pt][0pt]{\rm 0}}}
\put(105,730){\makebox(0,0)[lb]{\raisebox{0pt}[0pt][0pt]{\rm 2}}}
\put(205,765){\makebox(0,0)[lb]{\raisebox{0pt}[0pt][0pt]{\rm 0}}}
\put( 30,760){\makebox(0,0)[lb]{\raisebox{0pt}[0pt][0pt]{$(1^7)$}}}
\end{picture}.}

\medskip

The reader can verify that these maps are indeed reversible,
$\rho\circ\tau=\tau\circ\rho=$ id, and we shall refer to
the bijections thus described as {\kkr}.

It is important to note that the word $u$ gives rise to an admissible
rigged configuration if and only if it is a lattice permutation.

\begin{pr}
A rigged configuration is admissible
if and only if the multiset permutation of integers $0,\ldots,(n-1)$
in bijective correspondence with it is a lattice permutation.
\end{pr}
\noindent {\em Proof:} We shall prove that if we have an admissible
configuration, the associated word which is the sequence of ranks
read off in succession by ramification is a lattice permutation.
Since \kkr $\,$ is a bijection, the only if part follows.
In fact,
we shall prove that if the array encoding the word is not of partition
shape, {\em i.e.} $\lambda_a<\lambda_{a+1}$, then the configuration
is not admissible, {\em i.e.,} in particular, $h_{l_*}^{(a)}<0$,
where $l_*$ is the longest length of string in $\nu^{(a)}$.

For any colour $b$, $Q_{l_*}^{(b)}\leq |\nu^{(b)}|$, and
$Q_{l_*}^{(a)}=|\nu^{(a)}|$.  Therefore,
$Q_{l_*}^{(a-1)}-Q_{l_*}^{(a)}\leq \lambda_a$ and
$Q_{l_*}^{(a)}-Q_{l_*}^{(a+1)}\geq \lambda_{a+1}$.  Thus,
$$h_{l_*}^{(a)}=(Q_{l_*}^{(a-1)}-Q_{l_*}^{(a)})-
(Q_{l_*}^{(a)}-Q_{l_*}^{(a+1)})\leq (\lambda_a-\lambda_{a+1})<0.$$

\hfill{$\Box$}


\section{Strings and Holes: A pictorial construction}

The picture of strings and holes is vital to the physical interpretation
of the Bethe ansatz computations.  Key physical ideas like the notion
of ``filling the Dirac sea'', ``hole-excitations'', ``edge of the
Brillouin zone'' etc., are conveniently expressed in pictorial terms.
In this section, we shall present a pictorial representation of the
rigged configurations which is closer to this picture of strings and
holes.  It becomes easy to reformulate the ``ramification rule'' and
its inverse (the $\tau$ and $\rho$ maps of the \kkr $\,$ bijection) in
pictorial terms.  This reformulation is presented in \cite{Han}.

Let us focus on strings of any particular colour $a$ and length $l$.
Draw $s_l^{(a)}$ crosses and $h_l^{(a)}$ circles in a line
to represent the strings and holes and their relative positions
are determined by the integer riggings
$J_{\alpha},\alpha=1,\ldots,s_l^{(a)}$. The number of holes
($\circ$'s) to the {\em left}
of a particular string ($\times$) is the
rigging integer $J_{j,\mu}^{(a)}$.

\leftline{\bf Example.}

\centerline{
\setlength{\unitlength}{0.00725in}
\begin{picture}(331,135)(0,0)
\drawline(295,20)(295,0)(315,0)(315,20)
\drawline(295,40)(295,20)(315,20)(315,40)
\drawline(295,60)(295,40)(315,40)(315,60)
\drawline(295,80)(295,60)(315,60)(315,80)
\drawline(295,100)(295,80)(315,80)(315,100)
\drawline(295,120)(295,100)(315,100)(315,120)
\drawline(295,120)(315,120)
\put(0,40){\makebox(0,0)[lb]{\raisebox{0pt}[0pt][0pt]{\shortstack[l]{{$ \ex \ex
\oo \ex \oo \oo \ex \ex \ex \oo$}}}}}
\put(180,40){\makebox(0,0)[lb]{\raisebox{0pt}[0pt][0pt]{\shortstack[l]{{
$\Longleftrightarrow$}}}}}
\put(300,105){\makebox(0,0)[lb]{\raisebox{0pt}[0pt][0pt]{\shortstack[l]{$
\! \mbox{0}$}}}}
\put(300,85){\makebox(0,0)[lb]{\raisebox{0pt}[0pt][0pt]{\shortstack[l]{$
\! \mbox{0}$}}}}
\put(300,65){\makebox(0,0)[lb]{\raisebox{0pt}[0pt][0pt]{\shortstack[l]{$
\! \mbox{1}$}}}}
\put(300,45){\makebox(0,0)[lb]{\raisebox{0pt}[0pt][0pt]{\shortstack[l]{$
\! \mbox{3}$}}}}
\put(300,25){\makebox(0,0)[lb]{\raisebox{0pt}[0pt][0pt]{\shortstack[l]{$
\! \mbox{3}$}}}}
\put(300,5){\makebox(0,0)[lb]{\raisebox{0pt}[0pt][0pt]{\shortstack[l]{$
\! \mbox{3}$}}}}
\put(325,50){\makebox(0,0)[lb]{\raisebox{0pt}[0pt][0pt]{\shortstack[l]{\mbox{
4}}}}}
\end{picture}
}


This translation can be effected on the complete rigged
configuration where each linear distribution of crosses and
circles are placed in a rectangular grid, with the colour
and length indices increasing along the Cartesian $x$ and $y$
axes respectively, {\em i.e.} from left to right and from bottom
to top.  A line of strings and holes of a particular colour and
length is said to occupy a {\em zone}.  We shall denote by the
same symbol $\nu^{(a)}$, the set of zones of colour $a$.
We introduce an additional
set of zones in order to accommodate a set of {\em holes} of
colour $n$ such that the condition
$$\sum_a  a h_j^{(a)}=L$$
is satisfied.

\subsection{Ramification rule.}

Let us look at an example of the ramification rule of
the map $\tau$ in pictures.  For the two rigged configurations
shown in Figure \ref{bothfigs} (a), we draw the corresponding
configurations of
strings and holes in Figure \ref{bothfigs} (b) below.


\begin{figure}
\label{bothfigs}
\setlength{\unitlength}{0.00825in}
\begin{picture}(316,175)(0,-10)
\drawline(80,160)(80,0)(100,0)
	(100,120)(120,120)(120,140)
	(160,140)(160,160)(80,160)
\drawline(220,160)(220,120)(260,120)
	(260,140)(300,140)(300,160)(220,160)
\drawline(80,20)(100,20)
\drawline(80,40)(100,40)
\drawline(80,60)(100,60)
\drawline(80,80)(100,80)
\drawline(80,100)(100,100)
\drawline(80,120)(100,120)
\drawline(80,140)(120,140)
\drawline(100,160)(100,120)
\drawline(120,160)(120,140)
\drawline(140,160)(140,140)
\drawline(220,140)(260,140)(260,160)
\drawline(240,160)(240,120)
\drawline(280,160)(280,140)
\put(0,80){\makebox(0,0)[lb]{\raisebox{0pt}[0pt][0pt]{\shortstack[l]{{$(1^{18})$}}}}}
\put(85,5){\makebox(0,0)[lb]{\raisebox{0pt}[0pt][0pt]{\shortstack[l]{{\rm
3}}}}}
\put(85,25){\makebox(0,0)[lb]{\raisebox{0pt}[0pt][0pt]{\shortstack[l]{{\rm
3}}}}}
\put(85,45){\makebox(0,0)[lb]{\raisebox{0pt}[0pt][0pt]{\shortstack[l]{{\rm
3}}}}}
\put(85,85){\makebox(0,0)[lb]{\raisebox{0pt}[0pt][0pt]{\shortstack[l]{{\rm
0}}}}}
\put(85,105){\makebox(0,0)[lb]{\raisebox{0pt}[0pt][0pt]{\shortstack[l]{{\rm
0}}}}}
\put(85,125){\makebox(0,0)[lb]{\raisebox{0pt}[0pt][0pt]{\shortstack[l]{{\rm
0}}}}}
\put(85,145){\makebox(0,0)[lb]{\raisebox{0pt}[0pt][0pt]{\shortstack[l]{{\rm
0}}}}}
\put(225,145){\makebox(0,0)[lb]{\raisebox{0pt}[0pt][0pt]{\shortstack[l]{{\rm
0}}}}}
\put(225,125){\makebox(0,0)[lb]{\raisebox{0pt}[0pt][0pt]{\shortstack[l]{{\rm
2}}}}}
\put(270,125){\makebox(0,0)[lb]{\raisebox{0pt}[0pt][0pt]{\shortstack[l]{{\rm
2}}}}}
\put(310,145){\makebox(0,0)[lb]{\raisebox{0pt}[0pt][0pt]{\shortstack[l]{{\rm
0}}}}}
\put(165,145){\makebox(0,0)[lb]{\raisebox{0pt}[0pt][0pt]{\shortstack[l]{{\rm
0}}}}}
\put(125,125){\makebox(0,0)[lb]{\raisebox{0pt}[0pt][0pt]{\shortstack[l]{{\rm
2}}}}}
\put(110,55){\makebox(0,0)[lb]{\raisebox{0pt}[0pt][0pt]{\shortstack[l]{{\rm
4}}}}}
\put(85,65){\makebox(0,0)[lb]{\raisebox{0pt}[0pt][0pt]{\shortstack[l]{{\rm
1}}}}}
\end{picture}




\setlength{\unitlength}{0.00825in}
\begin{picture}(291,175)(0,-10)
\drawline(80,20)(100,20)
\drawline(80,40)(100,40)
\drawline(80,60)(100,60)
\drawline(80,80)(100,80)
\drawline(80,100)(100,100)
\drawline(80,120)(100,120)
\drawline(80,140)(120,140)
\drawline(100,160)(100,120)
\drawline(120,160)(120,140)
\drawline(220,140)(260,140)(260,160)
\drawline(240,160)(240,120)
\drawline(280,160)(280,140)
\drawline(140,160)(140,140)
\drawline(140,160)(80,160)(80,0)
	(100,0)(100,120)
\drawline(120,140)(140,140)
\drawline(100,120)(120,120)(120,140)
\drawline(220,120)(260,120)(260,140)(280,140)
\drawline(280,160)(220,160)(220,120)
\put(85,5){\makebox(0,0)[lb]{\raisebox{0pt}[0pt][0pt]{\shortstack[l]{{\rm
3}}}}}
\put(85,25){\makebox(0,0)[lb]{\raisebox{0pt}[0pt][0pt]{\shortstack[l]{{\rm
3}}}}}
\put(85,45){\makebox(0,0)[lb]{\raisebox{0pt}[0pt][0pt]{\shortstack[l]{{\rm
3}}}}}
\put(85,85){\makebox(0,0)[lb]{\raisebox{0pt}[0pt][0pt]{\shortstack[l]{{\rm
0}}}}}
\put(85,105){\makebox(0,0)[lb]{\raisebox{0pt}[0pt][0pt]{\shortstack[l]{{\rm
0}}}}}
\put(85,125){\makebox(0,0)[lb]{\raisebox{0pt}[0pt][0pt]{\shortstack[l]{{\rm
0}}}}}
\put(85,145){\makebox(0,0)[lb]{\raisebox{0pt}[0pt][0pt]{\shortstack[l]{{\rm
0}}}}}
\put(225,125){\makebox(0,0)[lb]{\raisebox{0pt}[0pt][0pt]{\shortstack[l]{{\rm
2}}}}}
\put(270,125){\makebox(0,0)[lb]{\raisebox{0pt}[0pt][0pt]{\shortstack[l]{{\rm
2}}}}}
\put(85,65){\makebox(0,0)[lb]{\raisebox{0pt}[0pt][0pt]{\shortstack[l]{{\rm
1}}}}}
\put(145,145){\makebox(0,0)[lb]{\raisebox{0pt}[0pt][0pt]{\shortstack[l]{{\rm
0}}}}}
\put(285,145){\makebox(0,0)[lb]{\raisebox{0pt}[0pt][0pt]{\shortstack[l]{{\rm
1}}}}}
\put(0,80){\makebox(0,0)[lb]{\raisebox{0pt}[0pt][0pt]{\shortstack[l]{{$(1^{17})$}}}}}
\put(110,55){\makebox(0,0)[lb]{\raisebox{0pt}[0pt][0pt]{\shortstack[l]{{\rm
3}}}}}
\put(125,125){\makebox(0,0)[lb]{\raisebox{0pt}[0pt][0pt]{\shortstack[l]{{\rm
1}}}}}
\put(225,145){\makebox(0,0)[lb]{\raisebox{0pt}[0pt][0pt]{\shortstack[l]{{\rm
1}}}}}
\end{picture}


\centerline{(a)}






\setlength{\unitlength}{0.00825in}
\begin{picture}(480,210)(0,-10)
\drawline(0,15)(70,15)(70,135)
	(330,135)(330,195)
\drawline(332.000,187.000)(330.000,195.000)(328.000,187.000)
\dashline{3.000}(480,160)(5,160)(5,0)
	(480,0)(480,160)
\put(85,20){\makebox(0,0)[lb]{\raisebox{0pt}[0pt][0pt]{\shortstack[l]{{$
\!\!\! \ex \ex \oo \ex \oo \oo \ex \ex \ex \, \oo $}}}}}
\put(100,60){\makebox(0,0)[lb]{\raisebox{0pt}[0pt][0pt]{\shortstack[l]{{$ \ex
\oo \oo$}}}}}
\put(105,100){\makebox(0,0)[lb]{\raisebox{0pt}[0pt][0pt]{\shortstack[l]{{$
\oo$}}}}}
\put(105,140){\makebox(0,0)[lb]{\raisebox{0pt}[0pt][0pt]{\shortstack[l]{{$
\ex$}}}}}
\put(275,20){\makebox(0,0)[lb]{\raisebox{0pt}[0pt][0pt]{\shortstack[l]{{$ \oo
\oo \oo  \, \oo $}}}}}
\put(280,60){\makebox(0,0)[lb]{\raisebox{0pt}[0pt][0pt]{\shortstack[l]{{$ \oo
\oo \ex$}}}}}
\put(285,100){\makebox(0,0)[lb]{\raisebox{0pt}[0pt][0pt]{\shortstack[l]{{$
\oo$}}}}}
\put(285,140){\makebox(0,0)[lb]{\raisebox{0pt}[0pt][0pt]{\shortstack[l]{{$
\ex$}}}}}
\put(395,20){\makebox(0,0)[lb]{\raisebox{0pt}[0pt][0pt]{\shortstack[l]{{$
 \oo \, \oo $}}}}}
\put(390,60){\makebox(0,0)[lb]{\raisebox{0pt}[0pt][0pt]{\shortstack[l]{{$
\oo \oo \oo \, \oo $}}}}}
\put(385,100){\makebox(0,0)[lb]{\raisebox{0pt}[0pt][0pt]{\shortstack[l]{{$
 \oo \oo \oo \oo \oo $}}}}}
\put(380,140){\makebox(0,0)[lb]{\raisebox{0pt}[0pt][0pt]{\shortstack[l]{{$
 \oo \oo \oo \oo \oo \, \oo $}}}}}
\put(15,20){\makebox(0,0)[lb]{\raisebox{0pt}[0pt][0pt]{\shortstack[l]{{$
\ex^{18}$}}}}}
\end{picture}





\setlength{\unitlength}{0.00825in}
\begin{picture}(480,185)(0,-10)
\dashline{3.000}(480,160)(5,160)(5,0)
	(480,0)(480,160)
\drawline(0,10)(260,10)(260,45)
	(360,45)(360,170)
\drawline(362.000,162.000)(360.000,170.000)(358.000,162.000)
\put(275,20){\makebox(0,0)[lb]{\raisebox{0pt}[0pt][0pt]{\shortstack[l]{{$ \oo
\oo \oo \, \oo$}}}}}
\put(280,60){\makebox(0,0)[lb]{\raisebox{0pt}[0pt][0pt]{\shortstack[l]{{$ \oo
\oo \ex$}}}}}
\put(395,20){\makebox(0,0)[lb]{\raisebox{0pt}[0pt][0pt]{\shortstack[l]{{$
\oo \, \oo$}}}}}
\put(390,60){\makebox(0,0)[lb]{\raisebox{0pt}[0pt][0pt]{\shortstack[l]{{$ \oo
\oo \oo \, \oo$}}}}}
\put(385,100){\makebox(0,0)[lb]{\raisebox{0pt}[0pt][0pt]{\shortstack[l]{{$ \oo
\oo \oo \oo \oo$}}}}}
\put(85,20){\makebox(0,0)[lb]{\raisebox{0pt}[0pt][0pt]{\shortstack[l]{{$ \ex
\ex \oo \ex \oo \oo \ex \ex \ex $}}}}}
\put(100,60){\makebox(0,0)[lb]{\raisebox{0pt}[0pt][0pt]{\shortstack[l]{{$ \ex
\oo $}}}}}
\put(105,100){\makebox(0,0)[lb]{\raisebox{0pt}[0pt][0pt]{\shortstack[l]{{$
\ex$}}}}}
\put(105,140){\makebox(0,0)[lb]{\raisebox{0pt}[0pt][0pt]{\shortstack[l]{{$
\cdot$}}}}}
\put(285,100){\makebox(0,0)[lb]{\raisebox{0pt}[0pt][0pt]{\shortstack[l]{{$ \oo
\ex$}}}}}
\put(285,140){\makebox(0,0)[lb]{\raisebox{0pt}[0pt][0pt]{\shortstack[l]{{$
\oo$}}}}}
\put(380,140){\makebox(0,0)[lb]{\raisebox{0pt}[0pt][0pt]{\shortstack[l]{{$ \oo
\oo \oo \oo \oo $}}}}}
\put(15,20){\makebox(0,0)[lb]{\raisebox{0pt}[0pt][0pt]{\shortstack[l]{{$
\ex^{17}$}}}}}
\end{picture}


\centerline{(b)}


\caption{\sf (a) The rigged configurations and (b) the string-hole
configurations for L=18 and L=17.}
\end{figure}

Note that the line drawn through the configuration, called the
{\em rank path} has the following properties.

\begin{alg}[Rank path]
The rank path is drawn {\em between} the zones of a configuration of
strings and holes, and the rules for navigating the zones are
such that
\begin{enumerate}
\item it starts at the {\em bottom left} and
\item it moves {\em under} a zone and to the {\em right}
if the {\em last} letter in that zone is a $\times$, and
\item it goes {\em in front of} a zone and {\em upwards}
if the last letter of that zone is a $\circ$.
\end{enumerate}
\end{alg}
\noindent Note that the tail of the rank path points upwards and out
of the configuration.

{}From the figure above, we can abstract the two basic ingredients of the
ramification rule that goes into the construction of the map $\tau$
from a configuration of strings and holes to a lattice permutation
which is canonically associated to a Young tableau.  These are:

\begin{alg}[Ramification rule]
For any two arbitrary distributions of $\times$ and $\circ$,
$w$ and $w'$ in neighbouring zones as depicted below, the last
$\times$ and $\circ$ are displaced according to the rules enumerated
below:
\begin{enumerate}
\item for all zones except those of length $1$ strings,
$$\begin{array}{c}
w\times \\
\longrightarrow  \\
w'
\end{array}
\Longrightarrow
\begin{array}{c}
w \\
\longrightarrow  \\
w' \times
\end{array}$$
\item for all zones except those of colour $1$ strings,
$$w \uparrow w'\circ \Longrightarrow w\circ \uparrow w'$$
\item for strings of length $1$,
$$\begin{array}{c}
w\times \\
\longrightarrow
\end{array}
\Longrightarrow
\begin{array}{c}
w \\
\longrightarrow
\end{array}$$
\item for zones of colours $0$ and $1$, if $x$ denotes a collection
of $\times$'s
$$x \uparrow w\circ \Longrightarrow x \uparrow w$$.
\end{enumerate}
\end{alg}

Using the rules enumerated, the next configuration after ramification
at length $L=16$ is given in Figure 3.  The reader can draw in the
rank path and verfy that the rank is $1$.  The Young tableau to which the
configuration corresponds to is
$$\begin{array}{cccccc}
1 & 3 & 5 & 6 & 10 & 13 \\
2 & 4 & 7 & 9 & 14 & 16 \\
8 & 11 & 12 & 15 & 17 & 18
\end{array}.$$

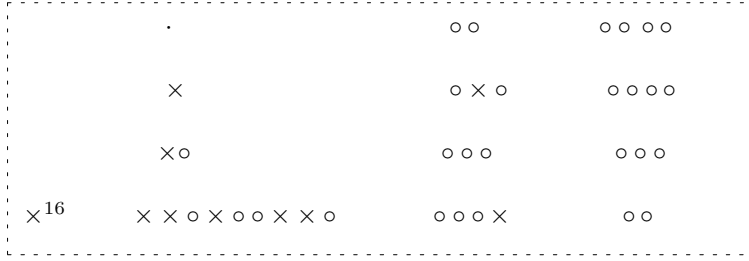
\begin{figure}
\setlength{\unitlength}{0.00825in}
\begin{picture}(480,185)(0,-10)
\dashline{3.000}(480,160)(5,160)(5,0)
	(480,0)(480,160)
\put(275,20){\makebox(0,0)[lb]{\raisebox{0pt}[0pt][0pt]{\shortstack[l]{{$ \oo
\oo \oo \, \ex$}}}}}
\put(280,60){\makebox(0,0)[lb]{\raisebox{0pt}[0pt][0pt]{\shortstack[l]{{$ \oo
\oo \oo$}}}}}
\put(395,20){\makebox(0,0)[lb]{\raisebox{0pt}[0pt][0pt]{\shortstack[l]{{$
\oo \, \oo$}}}}}
\put(390,60){\makebox(0,0)[lb]{\raisebox{0pt}[0pt][0pt]{\shortstack[l]{{$ \oo
\oo \oo $}}}}}
\put(385,100){\makebox(0,0)[lb]{\raisebox{0pt}[0pt][0pt]{\shortstack[l]{{$ \oo
\oo \oo\, \oo$}}}}}
\put(85,20){\makebox(0,0)[lb]{\raisebox{0pt}[0pt][0pt]{\shortstack[l]{{$ \ex
\ex \oo \ex \oo \oo \ex \ex \oo $}}}}}
\put(100,60){\makebox(0,0)[lb]{\raisebox{0pt}[0pt][0pt]{\shortstack[l]{{$ \ex
\oo $}}}}}
\put(105,100){\makebox(0,0)[lb]{\raisebox{0pt}[0pt][0pt]{\shortstack[l]{{$
\ex$}}}}}
\put(105,140){\makebox(0,0)[lb]{\raisebox{0pt}[0pt][0pt]{\shortstack[l]{{$
\cdot$}}}}}
\put(285,100){\makebox(0,0)[lb]{\raisebox{0pt}[0pt][0pt]{\shortstack[l]{{$ \oo
\ex \oo$}}}}}
\put(285,140){\makebox(0,0)[lb]{\raisebox{0pt}[0pt][0pt]{\shortstack[l]{{$ \oo
\, \oo$}}}}}
\put(380,140){\makebox(0,0)[lb]{\raisebox{0pt}[0pt][0pt]{\shortstack[l]{{$ \oo
\oo \,\oo \,\oo $}}}}}
\put(15,20){\makebox(0,0)[lb]{\raisebox{0pt}[0pt][0pt]{\shortstack[l]{{$
\ex^{16}$}}}}}
\end{picture}

\caption{\sf Ramification rules implemented, following on from Figure 2.
The rank of this configuration is $1$.}
\end{figure}

\section{Filtration property of the bijection}

In this section, we focus on rectangular tableaux, {\em i.e.}
those having the shape
$\lambda=(\lambda_1,\ldots,\lambda_n)$, with
$\lambda_1=\cdots=\lambda_n$, and note that the filtration of
paths by the level of the dominant weights translates into a filtration
of rigged
configurations of maximum allowed string lengths.  In terms of
Young diagrams, we have
$${\cal Y}_{n,1}\subset{\cal Y}_{n,2}\subset\cdots\subset{\cal Y}_{n,\ell}.$$
If ${\cal R}_k$
denotes the configuration of strings and holes with the maximum
length $k$ of the strings, then we have
$${\cal R}_1\subset {\cal R}_2\subset \cdots \subset{\cal R}_{l_{max}}.$$
In order to see the relation between $l_1$ and $l_2$ in
$${\cal Y}_{n,l_1}\stackrel{\mbox{\kkr}}{\longleftrightarrow}
{\cal R}_{l_2},$$
we need to consider a few lemmas before proving the principal result
below.

\newtheorem{lem}{Lemma}

\begin{lem}
If there is a string of length $k$ in $\nu^{(a+1)}$, there
must be one of length at least $k$ in $\nu^{(a)}$.
\label{length-order}
\end{lem}

\noindent {\em Proof:}
Let $|\lambda|=n p$, $p\in {\bf Z}_{>0}$, so $|\nu^{(a)}|=(n-a) p$.
Let the length of the
longest string in $\nu^{(a)}$ be $k$.  Then $Q_k^{(a)}=|\nu^{(a)}|$.
Assume that $\nu^{(a+1)}$ has at least one string of length $>k$, so
$Q_k^{(a+1)}<|\nu^{(a+1)}|=(n-a-1)p$, and $Q_k^{(a)}-Q_k^{(a+1)}>p$.
Also, $Q_k^{(a-1)}\leq|\nu^{(a-1)}|$, so $Q_k^{(a-1)}-Q_k^{(a)}\leq p$.
Thus,
\begin{equation}
\label{hQ}
h_k^{(a)}=(Q_k^{(a-1)}-Q_k^{(a)})-(Q_k^{(a)}-Q_k^{(a+1)})<0,
\end{equation}
and so $\nu$ is not an admissible configuration.
{\hfill $\Box$}

\begin{lem}
If the longest string in any configuration $(\vec{\nu},\{J\})$
corresponding to
$|\lambda|=n p$, $p\in{\bf Z}$ is of length $l_{*}$, then
the longest row in $\nu^{(n-1)}$ is of length $l_{*}$ too.
\label{lemtwo}
\end{lem}

\noindent {\em Proof}: Let the maximum length of strings for
colour $a_*$ be $l_{*}$, and assume that the length of the
longest string for colour $a_*+1$ is $l'<l_{*}$. By (\ref{hQ}),
$h_{l_*}^{(a_*)}=0=h_{l_*}^{(a_*+1)}$, since by
Lemma \ref{length-order},
none of
the strings in $\nu^{(b)}, b>a_*+1$ can have string lengths greater
than $l'$.  However, if at any stage of ramification, the topmost
zone of $\nu^{(a_*)}$ were to be above the rank path, the highest zone of
$\nu^{(a_*+1)}$ is then empty, and
consequently undefined and illegal.
It cannot have a circle, since $h_{l_*}^{(a_*+1)}=0$, so it
must have a cross.  Similarly, for all $\nu^{(b)},b>a_*+1$.
{\hfill $\Box$}

\begin{pr}[String length and level restrictions]
For $\Lambda'\in P^+\bigl(n;(\ell-1)\bigr)$,
$\Lambda\in P^+(n;1)$, let a path $(\mu,\eta)=\eta\in
{\cal P}_\ell^L(\Lambda',\Lambda)$ with $\mu_0=\ell\Lambda_0$
be encoded as a (normal) standard
tableau $T_\eta$, which is then mapped into a rigged configuration
$(\vec{\nu},\{J\})_\eta$.
The longest string length allowed in $(\vec{\nu},\{J\})_\eta$
is $\ell$, and there are no holes
in the zones containing $\ell$-strings.
\end{pr}

\noindent {\em Proof:}  We shall prove this by induction.  It is easy
to see that it is true for $\ell=1$,
where only strings of length $1$ are produced.
Let the tableau corresponding
to a level $\ell$ path be $T$, (dropping the subscripts) with
$\lambda=(\lambda_1,
\ldots,\lambda_n)=$sh$T$, $\lambda_1-\lambda_n=0$, and
$|\lambda|=(L+n)$.  The shape $\lambda$ and the shape of the
subtableau of $T$ with $L$ nodes are both rectangular.
For some $k<L$,
the partial tableau $T^{(k)}$ of $T$ with entries
$1,\ldots,k$ has shape
$\lambda^k=(\lambda_1^k,\ldots,\lambda_n^k)$,
$\lambda_1^k-\lambda_n^k=\ell$.
Let the rigged configuration $\rho(T^{(k)})$ be such that after
the steps
$w_{k+1}\ldots w_L$, the configuration
 has the longest strings of length $\ell$
with no holes in the zone for strings of length $\ell$,
by induction.
Just as for $\ell=1$, the steps  $w_{L+1},\ldots w_{L+n}$,
merely add a set of strings of length $1$, and $\rho(T)$
too has no holes in the zone for the $\ell$-strings.
Modify the path $w\mapsto w'$, by setting $w'_{k+1}=0$, and
$w'_{i+1}=w_i$ for $i=k+1,\ldots,L$, and $w'_{L+j}=w_{L+j}$
for $j=2,\ldots,n$.  Let the tableau corresponding to $w'$ be $T'$.

At the $(k+1)^{\scriptstyle th}$ step, $w'_{k+1}=0$, so
a hole ($\circ$) is introduced
to the right of each zone in $\nu^{(1)}$.  This is the only
effect
on the configuration.  For the steps $k+2,\ldots,L+1$, of
the modified
sequence $w'$, the corresponding changes to the
configurations of
strings and holes is exactly as for
$w$.  However,
the tableau $T'$ is not rectangular $-$
the first
row is longer than all the others by $1$.  Now compare the
strings and holes
after $L+1$ steps for the two cases $w$ and $w'$.
Both configurations
have the same
number of $\times$'s and $\circ$'s, but they are differently
arranged in the zones for colour $(1)$.
Recall that in the zone for length $\ell$
strings,
there were {\em no} holes after $w_L=n-1$, by the induction
hypothesis, but after
$w_{L+1}=0$, there
is a hole ($\circ$) at the end.  Any rearrangement of this
distribution {\em must necessarily contain an edge string}.
Thus, after $w'_{L+1}=w_L=(n-1)$, the configuration
$\rho(T^{' (L+1)})$ contains an edge string of length $\ell$ and
colour $1$, which implies that with $w'_{L+2}=1$,
a string of length $\ell+1$ is produced.  By Lemma \ref{lemtwo},
strings of length $(\ell+1)$ are produced for all colours,
with no holes. This completes the
induction step.
{\hfill $\Box$}

\section{On weights: momenta of Bethe states}

In the standard machinery related to the Bethe ansatz \cite{Tak},
the momenta
and energies of the states are calculated from the data provided
by the strings/holes picture.  For every Bethe state, the momentum
is given by ${2\pi\over L}{\cal I}(\vec{\nu},\{J\})$, where
\begin{equation}
\label{I-sum}
{\cal
I}(\vec{\nu},\{J\})=\sum_{a=1}^{n-1}\sum_{j=1}^{\ell}\sum_{\mu=1}^{s_j^{(a)}}
\biggl(I_{j,\mu}^{(a)}=J_{j,\mu}^{(a)}+\mu-{1\over 2}(s_j^{(a)}+h_j^{(a)}+1)
\biggr).
\end{equation}
It is convenient to
decompose the sum into a piece which has all the $J^{(a)}_{\alpha}=0$
and a sum over the integers $J^{(a)}_{\alpha}$.
For regime III of the model, the appropriate variable for expressing the
momenta in order to be amenable for the calculation of the excitation
spectrum is the number of holes, $h_j^{(a)}$.  Using (\ref{s-h}), the
sum reduces to
\begin{equation}
\begin{array}{ccl}
{\cal I}(\vec{\nu},\{J\})&=&\displaystyle -{1\over 2}\sum_{a=1}^{n-1}
\sum_j s_j^{(a)}h_j^{(a)} +\sum_{b=1}^{n-1}\sum_{j=1}^\ell
\sum_{\alpha=1}^{s_j^{(b)}} J^{(b)}_{j,\alpha} \\
&=&\displaystyle
-{L\over 2}\sum_{a=1}^{n-1}G^{-1}_{a1}h_1^{(a)}
+{1\over 2}\sum_{ab\,;\,jk}G^{-1}_{ab}C_{jk}h_j^{(a)}h_k^{(b)}
+\sum_{b=1}^{n-1}\sum_{j=1}^\ell\sum_{\alpha=1}^{s_j^{(b)}}
J^{(b)}_{j,\alpha}.
\end{array}
\end{equation}

Our main concern in this section is to demonstrate the following
equality (Proposition \ref{II}):
\begin{equation}
{\cal I}(\vec{\nu},\{J\})=\mbox{\sf I}(T,L),
\end{equation}
from which we can infer properties of the energy of a path
in terms of the momentum of the state, or a configuration of
strings and holes, the two being in bijective correspondence.
We shall do this by an inductive argument, for which we need
to evaluate
$$\partial_L {\cal I}:={\cal I}(\vec{\nu},\{J\})-
{\cal I}(\widetilde{\vec{{\nu}},\{J\}})$$
where the configurations are related by ramification,
$|\nu^{(0)}|=L$ and $|\widetilde{{\nu}^{(0)}}|=L-1$.

{}From the algorithm described earlier, only the changes
to the zones on either side of the rank path have to be
considered.  The local changes in ${\cal I}$ across a purely
horizontal section,
$$\begin{array}{c}
w\times \\
{\longrightarrow}  \\
w'
\end{array}
\Longrightarrow
\begin{array}{c}
w \\
{\longrightarrow}  \\
w' \times
\end{array}$$
are
$$\begin{array}{rcl}
{\cal I}(w)&=&{\cal I}(w \times) - {1\over 2} h_{w\times}\\
{\cal I}(w' \times)&=&{\cal I}(w') + {1\over 2} h_{w'}.
\end{array}
$$
Similarly, the changes across a purely vertical rank path,
$$w \uparrow w'\circ \Longrightarrow w\circ \uparrow w'$$
are
$$\begin{array}{rcl}
{\cal I}(w \circ)&=&{\cal I}(w) - {1\over 2} s_{w}\\
{\cal I}(w')&=&{\cal I}(w' \circ) + {1\over 2} s_{w'\circ}.
\end{array}
$$

\def\concave{
\setlength{\unitlength}{0.00325in}
\begin{picture}(40,55)(0,-10)
\drawline(0,0)(0,40)(40,40)
\drawline(32.000,38.000)(40.000,40.000)(32.000,42.000)
\end{picture}
}
\def\convex{
\setlength{\unitlength}{0.00325in}
\begin{picture}(40,55)(0,-10)
\drawline(0,0)(40,0)(40,40)
\drawline(42.000,32.000)(40.000,40.000)(38.000,32.000)
\end{picture}
}
At a convex corner, \hspace{-0.1in}\raisebox{-.6ex}{\convex},
$w\times\mapsto w\circ$, while at a concave corner,
\hspace{-0.1in}\raisebox{-.6ex}{\concave},
$w\circ\mapsto w\times$.  The changes in ${\cal I}$ are:
$$ \begin{array}{rcl}
{\cal I}(w \circ)&=&{\cal I}(w\times) - {1\over 2}
(s_{w\times}+h_{w\times}-1)\\
{\cal I}(w'\times)&=&{\cal I}(w' \circ)+{1\over 2} (s_{w'\circ}+h_{w'\circ}-1).
\end{array}
$$

Thus, if we represent the sum of integers in two neighbouring zones
by bracketing the two zones together, we can write
the local change in $\partial{\cal I}$ across a
horizontal section as follows.
$$\partial {\cal I}\left(\begin{array}{c}
{j+1} \\
{\longrightarrow}  \\
j
\end{array}\right)={\cal I}
\left(\begin{array}{c}
w{\times}_{j+1} \\
{\longrightarrow}  \\
w'_j
\end{array}\right)
- {\cal I}\left(\begin{array}{c}
w_{j+1} \\
{\longrightarrow}  \\
w' {\times}_j
\end{array}\right)={1\over 2}(h_j-h_{j+1}),$$
where $j$ and $(j+1)$ denote the string
lengths in the respective zones.  The corresponding
change in $\partial {\cal I}$ across a vertical section,
$$\partial {\cal I}\bigl(a \uparrow a+1\bigr)=
{\cal I}\bigl(w^{(a)}\uparrow w'\circ^{(a+1)}\bigr)-
{\cal I}\bigl(w \circ^{(a)} \uparrow {w'}^{(a+1)}\bigr)=
{1\over 2}(s^{(a+1)}-s^{(a)}),$$
where only the colour index has been indicated.
For convex and concave corners, both these effects add.

As before, for the part of the rank path below the strings of length $1$,
the rules are different.  But we shall return to that a little later.

%
%

\begin{lem}[Deformation of rank path]
Given a configuration of strings and holes and a rank path drawn
through it, perform any rearrangement of the strings and
holes such that the rank path needs to be deformed.
For any such deformation
the net change in ${\cal I}$
during ramification unchanged, unless the modification is such
that the strings of colour $1$ and length $1$
lie above the rank path in one case, and below it in the other.
In this latter instance, there is a net change of ${1\over 2}(L-1)$
for $L$ the length before ramification.
\label{deform}
\end{lem}

\noindent{\em Proof:}
Let the initial rank path end up at the value of the rank set at $b$.
Introduce a kink such that the modified rank path deviates from the
initial from string length $k$ and ends up at $b+1$. (See figure.)

\medskip

\medskip

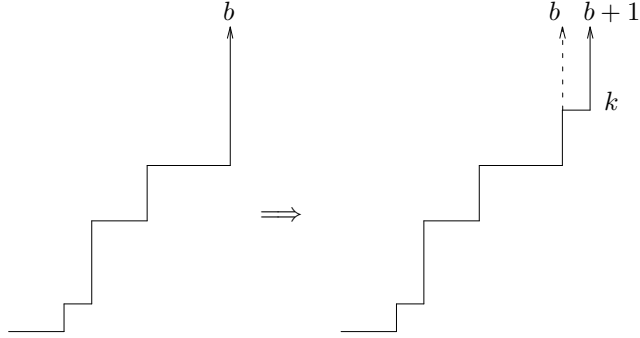
\centerline{
\setlength{\unitlength}{0.00725in}
\begin{picture}(446,235)(0,-10)
\drawline(0,0)(40,0)(40,20)
	(60,20)(60,80)(100,80)
	(100,120)(160,120)(160,220)
\drawline(162.000,212.000)(160.000,220.000)(158.000,212.000)
\drawline(240,0)(280,0)(280,20)
	(300,20)(300,80)(340,80)
	(340,120)(400,120)
\drawline(400,120)(400,160)(420,160)(420,220)
\drawline(422.000,212.000)(420.000,220.000)(418.000,212.000)
\dashline{4.000}(400,160)(400,220)
\drawline(402.000,212.000)(400.000,220.000)(398.000,212.000)
\put(175,80){\makebox(0,0)[lb]{\raisebox{0pt}[0pt][0pt]{\shortstack[l]{{
$\Longrightarrow$}}}}}
\put(155,225){\makebox(0,0)[lb]{\raisebox{0pt}[0pt][0pt]{\shortstack[l]{{$
b$}}}}}
\put(390,225){\makebox(0,0)[lb]{\raisebox{0pt}[0pt][0pt]{\shortstack[l]{{$
b$}}}}}
\put(415,225){\makebox(0,0)[lb]{\raisebox{0pt}[0pt][0pt]{\shortstack[l]{{$
b+1$}}}}}
\put(430,160){\makebox(0,0)[lb]{\raisebox{0pt}[0pt][0pt]{\shortstack[l]{{$
k$}}}}}
\end{picture}}

\noindent The only change this deformation induces in
$\partial {\cal I}$ is
$$-\, \, {1\over 2}\sum_{j\geq k} (s_j^{(b+1)}-s_j^{(b)}) \, \,
+\, \, {1\over 2}\sum_{j\geq k} (s_j^{(b+2)}-s_j^{(b+1)})+{1\over 2}
(h_{k-1}^{(b+1)}-h_{k}^{(b+1)}).$$
Using (\ref{bar-h}), this leads to the result that the net difference
in $\partial {\cal I}$ across the
two rank paths is zero.

However, if the rank line is modified so as to go under the zone for
stirngs of length $1$ and colour $1$, {\em i.e.}, $k=1, b=0$, there
is a residual $+{1\over 2}$ from the pairwise cancellations of the
factors of ${1\over 2}$ from $w\times\mapsto w\circ$
and $w\circ\mapsto w\times$, and note that only $h_1^{(1)}$ comes with
the additive piece, $L$.
{\hfill $\Box$}

\begin{lem}
The net change in ${\cal I}$ during a ramification for which the
length $L\mapsto L-1$ is:

\noindent (i) ${1\over 2}\bigl((L-1)-Q_1^{(1)}\bigr)$ if the rank path
goes below the zone for strings of length $1$, and

\noindent (ii) $-{1\over 2} Q_1^{(1)}$ otherwise.
\label{dI}
\end{lem}

\noindent {\em Proof}: Consider a configuration of rank $0$.  The change
in ${\cal I}$ across the (vertical) rank path is
$-{1\over 2}\sum_k s_k^{(1)}$.  There are two possibilities for modifying
the rank path, both covered by  Lemma \ref{deform}.  The results follow.
{\hfill $\Box$}

%
\begin{figure}
\setlength{\unitlength}{0.00825in}
\centerline{\begin{picture}(240,320)(0,-10)
\thicklines
\drawline(0,20)(60,20)(60,60)
	(120,60)(120,100)(140,100)
	(140,120)(200,120)(200,200)
	(220,200)(220,305)
\drawline(0,20)(60,20)(60,60)
	(120,60)(120,100)(140,100)
	(140,120)(200,120)(200,200)
	(220,200)(220,305)
\drawline(224.000,289.000)(220.000,305.000)(216.000,289.000)
\drawline(240,280)(240,0)(20,0)
	(20,280)(240,280)
\put(40,25){\makebox(0,0)[lb]{\raisebox{0pt}[0pt][0pt]{\shortstack[l]{{$\oo$}}}}}
\put(40,40){\makebox(0,0)[lb]{\raisebox{0pt}[0pt][0pt]{\shortstack[l]{{$\oo$}}}}}
\put(65,40){\makebox(0,0)[lb]{\raisebox{0pt}[0pt][0pt]{\shortstack[l]{{$\ex$}}}}}
\put(85,40){\makebox(0,0)[lb]{\raisebox{0pt}[0pt][0pt]{\shortstack[l]{{$\ex$}}}}}
\put(105,40){\makebox(0,0)[lb]{\raisebox{0pt}[0pt][0pt]{\shortstack[l]{{$\ex$}}}}}
\put(105,65){\makebox(0,0)[lb]{\raisebox{0pt}[0pt][0pt]{\shortstack[l]{{$\oo$}}}}}
\put(105,80){\makebox(0,0)[lb]{\raisebox{0pt}[0pt][0pt]{\shortstack[l]{{$\oo$}}}}}
\put(125,85){\makebox(0,0)[lb]{\raisebox{0pt}[0pt][0pt]{\shortstack[l]{{$\ex$}}}}}
\put(125,105){\makebox(0,0)[lb]{\raisebox{0pt}[0pt][0pt]{\shortstack[l]{{$\oo$}}}}}
\put(145,105){\makebox(0,0)[lb]{\raisebox{0pt}[0pt][0pt]{\shortstack[l]{{$\ex$}}}}}
\put(165,105){\makebox(0,0)[lb]{\raisebox{0pt}[0pt][0pt]{\shortstack[l]{{$\ex$}}}}}
\put(185,105){\makebox(0,0)[lb]{\raisebox{0pt}[0pt][0pt]{\shortstack[l]{{$\ex$}}}}}
\put(180,125){\makebox(0,0)[lb]{\raisebox{0pt}[0pt][0pt]{\shortstack[l]{{$\oo$}}}}}
\put(180,145){\makebox(0,0)[lb]{\raisebox{0pt}[0pt][0pt]{\shortstack[l]{{$\oo$}}}}}
\put(180,160){\makebox(0,0)[lb]{\raisebox{0pt}[0pt][0pt]{\shortstack[l]{{$\oo$}}}}}
\put(180,180){\makebox(0,0)[lb]{\raisebox{0pt}[0pt][0pt]{\shortstack[l]{{$\oo$}}}}}
\put(205,185){\makebox(0,0)[lb]{\raisebox{0pt}[0pt][0pt]{\shortstack[l]{{$\ex$}}}}}
\put(205,205){\makebox(0,0)[lb]{\raisebox{0pt}[0pt][0pt]{\shortstack[l]{{$\oo$}}}}}
\put(205,225){\makebox(0,0)[lb]{\raisebox{0pt}[0pt][0pt]{\shortstack[l]{{$\oo$}}}}}
\put(205,245){\makebox(0,0)[lb]{\raisebox{0pt}[0pt][0pt]{\shortstack[l]{{$\oo$}}}}}
\put(205,265){\makebox(0,0)[lb]{\raisebox{0pt}[0pt][0pt]{\shortstack[l]{{$\oo$}}}}}
\end{picture}}
\caption{\sf This figure shows the last entry ($\times$ or $\circ$)
in each zone on either side of the rank path {\em after} ramification.}
\label{rank-descent}
\end{figure}

\begin{lem}
For a standard tableau $T$ in bijective correspondence with a rigged
configuration,
$$d(T)=|\des(T)|=Q_1^{(1)}.$$
\label{dQ}
\end{lem}

\noindent {\em Proof}:  Note that $Q_1^{(1)}$ counts the number of
crosses ($\times$) in the zones for colour $1$ ({\em i.e.} $\nu^{(1)}$).
In Figure \ref{rank-descent}, a ramification is depicted where the lengths
before and after are $L$ and $L-1$, and where a $\times$ disappears
from the zone for strings of length $1$, {\em i.e.}, $Q_1^{(1)}$ is
reduced by $1$.  We have also indicated the last
letters ($\times$ or $\circ$) of each zone {\em after} the
ramification.
Observe that the $\circ$'s form a wall which prevents the rank path
from going beyond it.  The new rank must necessarily be {\em less}
than before.  In other words, $L-1$ is added to
${\des(T)}$.

{\hfill $\Box$}

\def\maximal{
\setlength{\unitlength}{0.00825in}
\begin{picture}(480,210)(0,-10)
\drawline(0,10)(260,10)(260,45)
	(360,45)(360,170)
\drawline(362.000,162.000)(360.000,170.000)(358.000,162.000)

\dashline{3.000}(480,160)(5,160)(5,0)
	(480,0)(480,160)

\put(15,20){\makebox(0,0)[lb]{\raisebox{0pt}[0pt][0pt]{\shortstack[l]{{$
\ex^{18}$}}}}}
\put(85,20){\makebox(0,0)[lb]{\raisebox{0pt}[0pt][0pt]{\shortstack[l]{{$
 \oo \oo \oo \oo \ex \ex \ex \ex \ex \ex $}}}}}
\put(100,60){\makebox(0,0)[lb]{\raisebox{0pt}[0pt][0pt]{\shortstack[l]{{$
\oo \oo \ex$}}}}}
\put(105,100){\makebox(0,0)[lb]{\raisebox{0pt}[0pt][0pt]{\shortstack[l]{{$
\oo$}}}}}
\put(105,140){\makebox(0,0)[lb]{\raisebox{0pt}[0pt][0pt]{\shortstack[l]{{$
\ex$}}}}}
\put(275,20){\makebox(0,0)[lb]{\raisebox{0pt}[0pt][0pt]{\shortstack[l]{{$ \oo
\oo \oo \, \oo$}}}}}
\put(280,60){\makebox(0,0)[lb]{\raisebox{0pt}[0pt][0pt]{\shortstack[l]{{$ \oo
\oo \ex$}}}}}
\put(285,100){\makebox(0,0)[lb]{\raisebox{0pt}[0pt][0pt]{\shortstack[l]{{$
\oo$}}}}}
\put(285,140){\makebox(0,0)[lb]{\raisebox{0pt}[0pt][0pt]{\shortstack[l]{{$
\ex$}}}}}
\put(395,20){\makebox(0,0)[lb]{\raisebox{0pt}[0pt][0pt]{\shortstack[l]{{$
\oo \, \oo $}}}}}
\put(390,60){\makebox(0,0)[lb]{\raisebox{0pt}[0pt][0pt]{\shortstack[l]{{$
\oo \oo \oo \, \oo $}}}}}
\put(385,100){\makebox(0,0)[lb]{\raisebox{0pt}[0pt][0pt]{\shortstack[l]{{$
\oo \oo \oo \oo \oo $}}}}}
\put(380,140){\makebox(0,0)[lb]{\raisebox{0pt}[0pt][0pt]{\shortstack[l]{{$
\oo \oo \oo \oo \oo \, \oo $}}}}}
\end{picture}
}

\begin{pr} For $T\leftrightarrow (\vec{\nu},\{J\})$, related by
\kkr $\,$ with $|sh \, T|=L$,
\begin{equation}
{\cal I}(\vec{\nu},\{J\})=\mbox{\sf I}(T,L)
\end{equation}
\label{II}
\end{pr}

\noindent {\em Proof}:  Let the entry $L$ be deleted from $T$ to give
the tableau $\tilde{T}$.  Then either $(L-1)$ is in a higher row in
$T$ than $L$, in which case $\des(\tilde{T})=\des (T)\setminus\{L-1\}$,
or $\des(\tilde{T})=\des (T)$ otherwise.  In the first case,
{\sf I}$(T,L)-${\sf I}$(\tilde{T},L-1)={1\over2}(L-d(T)-1)$, whereas
for the latter the net change in {\sf I} is $-{1\over 2}d(T)$.
The result follows from Lemmas \ref{dI}, \ref{dQ}.
{\hfill $\Box$}

This result prompts us to define an involution on a rigged configuration
as follows.

\begin{de}[Involution]

Given an admissible rigged configuration $(\vec{\nu},\{J\})$,
we define another, $(\vec{\nu},\{J^\sigma\})$ by performing the
following operations (involutions)
$\sigma_{j,\alpha}^{(a)}$,
($1\leq a\leq (n-1)$;  $1\leq j\leq \ell$,
$1\leq\alpha\ldots,s_j^{(a)})$ and
$\sigma:=\prod_{a=1}^{n-1}\prod_{j=1}^\ell
\prod_{\alpha=1}^{s_j^{(a)}} \sigma_{j,\alpha}$:
\[
\label{J-inv}
\begin{array}{rcl}
\sigma_{j,\mu}^{(a)}:\{J\}&\longrightarrow &\{J\}\\
J_{j,\mu}^{(a)}&\mapsto &J^{\sigma\, (a)}_{j,\mu}=h_{j}^{(a)}-
J_{j,s^{(a)}_j+1-\mu}^{(a)}.
\end{array}
\]
Note that $\sigma_{j,\alpha}^{(a)\, 2}=\sigma^2=$id.
\end{de}

In pictorial terms, this involution is nothing but a reflection
about the centre of the line of noughts and crosses, so that
in terms of Takahashi integers,
$$\sigma:\{I\}\longrightarrow \{-I\}.$$

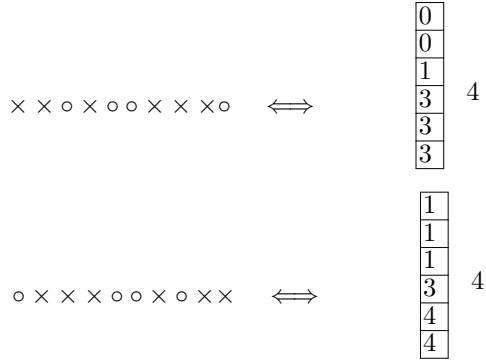
\begin{figure}
\setlength{\unitlength}{0.00725in}
\centerline{
\begin{picture}(331,135)(0,0)
\drawline(295,20)(295,0)(315,0)(315,20)
\drawline(295,40)(295,20)(315,20)(315,40)
\drawline(295,60)(295,40)(315,40)(315,60)
\drawline(295,80)(295,60)(315,60)(315,80)
\drawline(295,100)(295,80)(315,80)(315,100)
\drawline(295,120)(295,100)(315,100)(315,120)
\drawline(295,120)(315,120)
\put(0,40){\makebox(0,0)[lb]{\raisebox{0pt}[0pt][0pt]{\shortstack[l]{{$ \ex \ex
\oo \ex \oo \oo \ex \ex \ex \oo$}}}}}
\put(180,40){\makebox(0,0)[lb]{\raisebox{0pt}[0pt][0pt]{\shortstack[l]{{
$\Longleftrightarrow$}}}}}
\put(300,105){\makebox(0,0)[lb]{\raisebox{0pt}[0pt][0pt]{\shortstack[l]{$
\! \mbox{0}$}}}}
\put(300,85){\makebox(0,0)[lb]{\raisebox{0pt}[0pt][0pt]{\shortstack[l]{$
\! \mbox{0}$}}}}
\put(300,65){\makebox(0,0)[lb]{\raisebox{0pt}[0pt][0pt]{\shortstack[l]{$
\! \mbox{1}$}}}}
\put(300,45){\makebox(0,0)[lb]{\raisebox{0pt}[0pt][0pt]{\shortstack[l]{$
\! \mbox{3}$}}}}
\put(300,25){\makebox(0,0)[lb]{\raisebox{0pt}[0pt][0pt]{\shortstack[l]{$
\! \mbox{3}$}}}}
\put(300,5){\makebox(0,0)[lb]{\raisebox{0pt}[0pt][0pt]{\shortstack[l]{$
\! \mbox{3}$}}}}
\put(325,50){\makebox(0,0)[lb]{\raisebox{0pt}[0pt][0pt]{\shortstack[l]{\mbox{
4}}}}}
\end{picture}}

\centerline{
\setlength{\unitlength}{0.00725in}
\begin{picture}(331,135)(0,0)
\drawline(295,20)(295,0)(315,0)(315,20)
\drawline(295,40)(295,20)(315,20)(315,40)
\drawline(295,60)(295,40)(315,40)(315,60)
\drawline(295,80)(295,60)(315,60)(315,80)
\drawline(295,100)(295,80)(315,80)(315,100)
\drawline(295,120)(295,100)(315,100)(315,120)
\drawline(295,120)(315,120)
\put(0,40){\makebox(0,0)[lb]{\raisebox{0pt}[0pt][0pt]{\shortstack[l]{{$ \oo \ex
\ex \ex \oo \oo \ex \oo \ex \ex$}}}}}
\put(180,40){\makebox(0,0)[lb]{\raisebox{0pt}[0pt][0pt]{\shortstack[l]{{
$\Longleftrightarrow$}}}}}
\put(300,105){\makebox(0,0)[lb]{\raisebox{0pt}[0pt][0pt]{\shortstack[l]{$
\! \mbox{1}$}}}}
\put(300,85){\makebox(0,0)[lb]{\raisebox{0pt}[0pt][0pt]{\shortstack[l]{$
\! \mbox{1}$}}}}
\put(300,65){\makebox(0,0)[lb]{\raisebox{0pt}[0pt][0pt]{\shortstack[l]{$
\! \mbox{1}$}}}}
\put(300,45){\makebox(0,0)[lb]{\raisebox{0pt}[0pt][0pt]{\shortstack[l]{$
\! \mbox{3}$}}}}
\put(300,25){\makebox(0,0)[lb]{\raisebox{0pt}[0pt][0pt]{\shortstack[l]{$
\! \mbox{4}$}}}}
\put(300,5){\makebox(0,0)[lb]{\raisebox{0pt}[0pt][0pt]{\shortstack[l]{$
\! \mbox{4}$}}}}
\put(325,50){\makebox(0,0)[lb]{\raisebox{0pt}[0pt][0pt]{\shortstack[l]{\mbox{
4}}}}}
\end{picture}}
\caption{\sf An involution (reflection) acting on a rigged configuration.}
\label{inv-fig}
\end{figure}

\leftline{\bf Example.}  The action of the involution on the
(section of)  the rigged configuration above is depicted
in Figure \ref{inv-fig}.

\begin{pr}[\cite{KR}]
For $T,T^S\in SYT(\lambda)$ related by the evacuation involution,
the following diagram is commutative:
\[
\begin{array}{ccc}
T & \stackrel{S}{\longleftrightarrow} & T^S \\
\mbox{\kkr}\updownarrow & & \updownarrow\mbox{\kkr} \\
(\nu,\{J\}) & \stackrel{\sigma}{\longleftrightarrow} &
(\nu,\sigma\circ \{J\})
\end{array}
\]
\end{pr}

\noindent {\em Proof}: Follows from Propositions \ref{II} and
\ref{I-inv}.
{\hfill $\Box$}

\noindent {\bf Remark.}  The relation between
${\cal I}(\vec{\nu},\{J\})$ and $-{\cal I}(\vec{\nu},\{J\})$ is
one of computing momenta by strings or by holes.  Replacing a
string by a hole ($\times\leftrightarrow\circ$) and interchanging
rows and columns is the effect of implementing level-rank duality,
{\em i.e.} going from regime III to regime II.

\section{Polynomial identities and branching functions}

The results of the previous section allow us to make this section
rather short!  Recall, eq. (\ref{jmo-bosonic}) gives the $H$-weighted
sum over restricted paths in ${\cal P}_{\ell}(\Lambda, \Lambda')$.
We have demonstrated that these are in bijection with configurations
of strings and holes, and that the momentum of the Bethe states
can be used to compute the statistic {\bf p} of the corresponding
tableau, or, equivalently, $H$ of the corresponding path.

Putting everything together, we arrive at:
\begin{pr}
For $\Lambda'\in P^+\bigl(n;(\ell-1)\bigr)$,
$\Lambda\in P^+(n;1)$, let a path $(\mu,\eta)=\eta\in
{\cal P}_\ell^L(\Lambda',\Lambda)$ with $\mu_0=\ell\Lambda_0$
be encoded as a (normal) standard
tableau $T_\eta$, which is then mapped into a rigged configuration
$(\vec{\nu},\{J\})_\eta$.
Then,
\begin{equation}
H(\eta)={\bf p}(T_{\eta})={1\over2}L(L-Q_1^{(1)}-1)+\sum_{a=1}^{n-1}
\sum_{j=1}^{\ell}\sum_{\mu=1}^{s_j^{(a)}} I_{j \mu}^{(a)}
\end{equation}
\end{pr}

\noindent {\em Proof}: Follows from the definition of {\bf p}(T)
(eq. \ref{cp-defs}) and proposition \ref{II}. {\hfill $\Box$}

\noindent The corresponding generating functions are therefore
equal:

\begin{pr}[polynomial identity]
Let $\{\nu\}$ denote the set of rigged configurations
$(\vec{\nu},\{J\})$ obtained via the \kkr $\,$ map,
$(\vec{\nu},\{J\})=\rho(T)$, and $T$ be such that
$w_T=\eta$ such that $(\mu,\eta)\in{\cal P}^L_\ell\bigl((\ell-1)\Lambda_0,
\Lambda_0)$ with $\mu_0=\ell\Lambda_0$.
Also, set
$$({\bf h}|{\bf h})=\sum_{ab\,;\,jk}G^{-1}_{ab}C_{jk}h_j^{(a)}h_k^{(b)},$$
and
\begin{equation}
\label{myform}
F_L^{\ell}(q)=\sum_{\{\nu\}}
\prod_{a=1}^{n-1}\prod_{j=1}^{\ell}q^{{1\over 2}({\bf h}|{\bf h})}
\biggl[
\begin{array}{c}
h_j^{(a)}+\sum_b \sum_k G^{-1}_{ab}C_{jk}h_k^{(b)}\\
h_j^{(a)}
\end{array}\biggr].
\end{equation}
Then,
\begin{equation}
\sum_{T} q^{c(T)}=
\sum_{T} q^{\mbox{\boldmath p}(T)} =
q^{ {L\over 2}({L\over n}-1)}F_L^{\ell}(q) =
X_L(\ell\Lambda_0,\ell\Lambda_0,(\ell-1)\Lambda_0+\Lambda_1;q)
\end{equation}
\end{pr}

\noindent {\em Proof}:  Using eq. (\ref{s-h}), we get
\begin{equation}
Q_1^{(1)}=L G^{-1}_{11}-\sum_{a=1}^{n-1}G^{-1}_{1a}h_1^{(a)}.
\end{equation}
The result then follows from (\ref{I-sum}),
and the previous proposition. {\hfill $\Box$}

The excitation spectra are computed
by taking the deviations of the distribution of strings from the
ground state, following the early computations of \cite{Lieb}.
This requires the subtraction of the momenta of the holes from the
Fermi momentum, which determines the edge of the Brillouin zone for
the single particle states.  We thus need to know the ground state
configuration for the model in regime III, the case we are focusing
on for this paper.

For the ground state defined in the CTM language as the sequence
$(\mu^{\scriptstyle GS},\eta^{\scriptstyle GS})$
with $\mu_k^{\scriptstyle GS}=
(\ell-1)\Lambda_0+\Lambda_{k}$, the corresponding
rigged configuration is one in which there are strings of length
$1$ only.  For the case $\mu_L=\ell\Lambda_0$ ($L=nx$, integer $x$)
we have
\begin{equation}
s_j^{(a)}=\delta_{j1}(1-{a\over n})L, \qquad h_j^{(a)}={L\over n}
\delta_{an}\delta_{j1},
\end{equation}
and so for the zones of interest $a=1,\ldots,n-1$, there are
{\em no holes} in the ground state for regime III of the model.
This is also the observation of reference \cite{BRtwo,das}.
It is convenient \cite{BRtwo,das} to write out everything in terms
of the number of holes $h_j^{(a)}$, and we therefore have to
take the following sum in order to
get the momentum of the excited states, $\vec{P}$,
after subtracting off the Fermi momentum:
\begin{equation}
\label{quasi-mom}
\vec{P}(\vec{\nu},\{J\})={\cal I}(\vec{\nu},\{J\})-
{1\over 2}\sum_{a=1}^{n-1}\sum_{j=1}^{\ell}\sum_{\mu=1}^{h_j^{(a)}}
LG^{-1}_{a1}\delta_{j1}.
\end{equation}

Working this out, we find that both descriptions, setting a suitable
zero of the excitation momentum as well as subtracting off the energy
contribution of the ground state CTM path (recall the definition
of the energy $H(p)$ of a path $p$), gives rise to the same term to
be subtracted, namely $L\bigl((L/n)-1\bigr)/2$.
This leads to the proof of Theorem 1, as mentioned in Section 1 (the
introduction).

\section{Discussion}

For the polynomials corresponding to paths other than those
for which $\mu_0=\mu_L=\ell\Lambda_0$, we note that \kkr $\,$
gives an algorithm for obtaining the fermionic form for
$X_L\bigl(\ell\Lambda_0,\Lambda',\Lambda'+(\Lambda_{i+1}-\Lambda_i)\bigr)$.
The shape of the tableau is determined by $\Lambda'$.  As for
$\mu_0\not=\ell\Lambda_0$, we have to consider skew tableau,
of shape (say) $\lambda/\nu$.  Recall that for the unrestricted
(or $\ell\rightarrow\infty$) case, the Kostka-Foulkes
polynomials for skew and normal shapes are related by the
Littlewood-Richardson coefficients by the following argument.

For $s_{\tau/\nu}(X)$, the Schur function indexed by the skew
diagram $\tau/\nu$, we note that it may be expressed as the
linear combination $s_{\tau/\nu}(X)=\sum_{\mu}c_{\nu\mu}^{\tau}s_{\mu}(X)$.
$c_{\nu\mu}^{\tau}$ is the Littlewood-Richardson coefficient,
which (equivalently) describes the decomposition
$\Gamma_{\nu}\otimes \Gamma_{\mu}=\oplus c_{\nu \mu}^{\tau}
\Gamma_{\tau}$ of $gl(n)$ highest weight modules indexed by partitions,
given by \cite{Macdonald}
$$c_{\nu\mu}^{\tau}=|\{T|\mbox{sh} T=\tau/\nu, \mbox{wt}T =\mu,
\mbox{charge} T=0\}|.$$
Therefore,
$$s_{\tau\nu}(X)=\sum_{\mu}c_{\nu\mu}^{\tau}\bigl(\sum_{\lambda}
K_{\mu\lambda}(q) P_{\lambda}(X;q)\bigr),$$
so that if we define skew Kostka polynomials by
$$s_{\tau\nu}(X)=\sum_{\lambda}K_{\tau/\nu \, \lambda}(q)
P_{\lambda}(X;q),$$
we obtain
$$K_{\tau/\nu \, \lambda}(q)=\sum_{\mu}c_{\nu\mu}^{\tau}
K_{\mu\lambda}(q).$$

The combinatorial definition for skew tableaux in terms of {\bf p}
and the charge is analogous to that for normal shapes \cite{Butler},
so the
energy function is faithfully evaluated by the statistics on
skew tableaux.
The only difference for the {\em restricted} case is that the
Littlewood-Richardson coefficients are replaced by the fusion
rules \cite{verlinde}.  However, for the paths described by Young diagrams
${\cal Y}_{n,\ell}$, the two sets of coefficients are the same
\cite{GW}.

We hope to use this data in a future publication
to obtain explicit fermionic forms for
the other polynomials which give the rest of the branching functions
for the cosets under consideration. Also, for fusion models, we expect
that the proofs should follow in a straightforward manner, similar
to those in this paper.

The action of the vertex operators on the path basis can be lifted
over to the string-hole configurations by the bijection described.
This must surely be the most exciting prospect that has opened up.

\section{Acknowledgements}

I would like to thank Omar Foda for discussions,
and collaboration; this work is a direct extension of
ideas worked out in Melbourne while I was visiting University
of Melbourne.  I would also like to thank him for his hospitality
as well.  I would also like to acknowledge useful discussions
with Kailash Misra, Masato Okado and Ole Warnaar.  Special thanks to
Babak Razzaghe-Ashrafi for getting me the references \cite{KKR,KR}.
I would like to thank the EPSRC for financial support in the
form of the grants GRJ25758 and GRJ29923.

\end{document}